\documentstyle[preprint,aps,eqsecnum]{revtex}

\begin{document}
\tighten
\draft
\preprint{
 \parbox[t]{50mm}{hep-ph/9511403\\       %archives
                  DPNU-95-37     \\      %Nagoya
                  TUIMP-TH-95/71  \\ }}  %Peking

\title{Four-Body Chiral Order Parameters
       in the Massless Fermion Phase}

\author{Pieter Maris}
\address{Department of Physics,
         Nagoya University, Nagoya 464-01, Japan}
\author{Qing Wang}
\address{Department of Physics,
         Tsinghua University, Beijing 100084, China}

\date{November 15, 1995}
\maketitle
\begin{abstract}
The fermion four-point functions and condensates as the chiral
symmetry order parameters are calculated analytically in U(1) gauge
theory in the massless phase. It is shown that in leading order of the
loop expansion of the effective action, there is a critical coupling
for the nonperturbative parity-invariant chirality-changing
four-fermion functions; this critical coupling however is the same as
that for dynamical mass generation. Therefore there is no breaking of
chiral symmetry through the condensation of four-fermion functions
below the critical coupling for dynamical mass generation for the
fermions.
\end{abstract}
%

%%%%%%%%%%%%%%%%%%%%%%%%%%%%%%%%%%%%%%%%%%%%%%%%%%%%%%%%%%%%%%%%%%%%%%%%%%%%%
\section{Introduction}
%%%%%%%%%%%%%%%%%%%%%%%%%%%%%%%%%%%%%%%%%%%%%%%%%%%%%%%%%%%%%%%%%%%%%%%%%%%%%

Dynamical chiral symmetry breaking (DCSB) plays a very important role
in the research of fundamental interactions. Up to now, most
investigations are focussed on DCSB with mass generation for the
fermions. In such a scenario, the bilinear fermion condensate is taken
as the order parameter of chiral symmetry breaking and it is known
that it can receive a nonzero value when the coupling of the system
exceeds some critical value. For a review on dynamical chiral symmetry
breaking, see e.g.~\cite{MIR}.

But this is not the complete story for DCSB, since the order parameter
may not be the bilinear fermion condensate: in principle, higher
$n$-point functions which break chiral symmetry might become nonzero,
without dynamical mass generation for the fermions
\cite{HolTri951,HolTri95N}. In a massless $U(1)$ gauge theory for
example, four-fermion Green's functions which break chiral symmetry
might become nonzero at values of the coupling below the critical
value for dynamical mass generation. Such a breaking of the chiral
symmetry by the four-fermion condensates, without mass generation for
the fermions, was discussed recently by Holdom and Triantaphyllou
\cite{HolTri951}, who have calculated these four-point functions
numerically and found a critical coupling constant which is almost the
same as that responsible for dynamical mass generation.

One could also think of theories where some other symmetry prevents
the dynamical generation of a bilinear condensate and a dynamical
fermion mass, but where a nonzero condensate of a higher $n$-point
function is not restricted by this symmetry. In a chiral gauge theory,
the gauge symmetry will be broken by the generation of a fermion mass,
but a chirality-changing four-fermion condensate does not break the
gauge symmetry, and might therefore be generated dynamically. In such
a scenario, the chiral symmetry can be realized in more delicate way
which may be useful for model building \cite{YaYo82,Hol94,HolKyoto}:
it allows for a hierarchy of chiral symmetry breaking, first by the
four-fermion condensate at relatively small values of the coupling,
and eventually a second phase transition, at higher values of the
coupling, leading to mass generation for the fermions, due to the
four-fermion condensates. For these purposes one should study this
phenomenon in a chiral gauge theory, as suggested in \cite{HolKyoto},
but for simplicity and to see whether such a symmetry breaking
patterns is possible, we restrict ourselves to a $U(1)$ gauge theory.

In our paper we confirm analytically the numerical results obtained by
Holdom and Triantaphyllou \cite{HolTri951}, using a slightly different
method to calculate the nonperturbative four-fermion functions. We
find that there is a critical coupling for dynamical chiral symmetry
breaking via the four-fermion condensates, below which there are no
nontrivial solutions for the four-fermion functions which break chiral
symmetry, but preserve parity, Lorentz symmetry, and the flavor
symmetry. This critical coupling is (in our truncation scheme) exactly
the same as the critical coupling for fermion mass generation. This
result is obtained by calculating the loop expansion of the effective
action to leading order for the nonperturbative four-point functions.
Our result is also in agreement with an intuitive argument comparing
the condensation of two fermions to the condensation of four fermions:
for the four-fermion condensate we need at least the same binding
force as for the bilinear fermion condensate.

Our paper is organized as follows. In the next section, we discuss
some general features about dynamical symmetry breaking, and we show
how in principle the chiral symmetry can be broken without mass
generation for the fermions. In Sec.~\ref{seceffact}, we derive a
linear integral equation for the four-fermion function, using the loop
expansion of the effective action. Next, we analyze this equation for
the nonperturbative chirality-changing four-fermion functions, using
some expansions in order to perform the angular integrations
analytically; more details can be found in the appendix. We find that
there is a critical coupling for these functions; this critical
coupling turns out to be the same as the critical coupling for
dynamical mass generation. Finally we give some concluding remarks in
Sec.~\ref{secconc}.

%%%%%%%%%%%%%%%%%%%%%%%%%%%%%%%%%%%%%%%%%%%%%%%%%%%%%%%%%%%%%%%%%%%%%%%%%%%%%
\section{General Considerations on Dynamical Symmetry Breaking}
\label{secgeneralsym}
%%%%%%%%%%%%%%%%%%%%%%%%%%%%%%%%%%%%%%%%%%%%%%%%%%%%%%%%%%%%%%%%%%%%%%%%%%%%%

\subsection{Symmetry Breaking Patterns}

The kind of symmetry breaking we are dealing with in this paper is the
breaking of chiral symmetry in a $U(1)$ gauge theory with $N_f$
massless fermion flavors. The global continuous chiral symmetry is
$U_{L}(N_{f})\otimes U_{R}(N_{f})$, and this symmetry can be broken in
different ways, characterized by the corresponding chiral order
parameters. Since the flavor symmetry is not broken in vector-like
gauge theories \cite{VWI}, only flavor-symmetry invariant order
parameters may receive nonzero contributions; also parity \cite{VWP}
and Lorentz symmetry is believed not to be broken spontaneously, so in
this paper we limit us to parity-, flavor-, and Lorentz-invariant
order parameters.

The simplest and most wellknown way of breaking the chiral symmetry
is characterized by the trace of the two-point function
$\langle 0|{\bf T}{\bar\psi}(x)\psi(y)|0\rangle$. This condensate,
$\langle 0|{\bar\psi}(x)\psi(x)|0\rangle $, is the only possible
two-fermion condensate which keeps the flavor symmetry and is parity
and Lorentz invariant. It gives rise to a dynamically generated
fermion mass, and the symmetry breaking pattern is
\begin{eqnarray}
  U_{L}(N_{f})\otimes U_{R}(N_{f})
    & \longrightarrow & U_{L+R}(N_{f}) \,.
\label{0.1}
\end{eqnarray}
An explicit mass for the fermions would also break the symmetry in
this way.

However, this is not the only way to break the chiral symmetry, an
alternative way is characterized by condensates of four fermions. The
possible four-point function which keep flavor and Lorentz symmetry
and which are parity invariant are
\begin{eqnarray}
 {\cal G}_{0, S+P}(x_1,x_2,x_3,x_4) & = &
 \langle 0|{\bf T}{\bar\psi}(x_1)\lambda^{0}\psi (x_2)
  {\bar\psi}(x_3)\lambda^{0}\psi (x_4)|0\rangle
 \nonumber\\&& +
   \langle 0|{\bf T}{\bar\psi}(x_1)\gamma_{5}\lambda^{0}\psi (x_2)
  {\bar\psi}(x_3)\gamma_{5}\lambda^{0}\psi (x_4)|0\rangle  \,,
 \label{0.01}\\
 {\cal G}_{S+P}(x_1,x_2,x_3,x_4) & = &
 \sum_{\alpha = 1}^{N_f^2-1}\bigg[
 \langle 0|{\bf T}{\bar\psi}(x_1)\lambda^{\alpha}\psi (x_2)
  {\bar\psi}(x_3)\lambda^{\alpha}\psi (x_4)|0\rangle
 \nonumber\\&& +
  \langle 0|{\bf T}{\bar\psi}(x_1)\gamma_{5}\lambda^{\alpha}
  \psi (x_2){\bar\psi}(x_3)\gamma_{5}\lambda^{\alpha}\psi (x_4)
  |0\rangle \bigg] \,,
 \label{0.02}\\
 {\cal G}_{0,T}(x_1,x_2,x_3,x_4) & = &
  \langle 0|{\bf T}{\bar\psi}(x_1)\sigma_{\mu\nu}\lambda^{0}
  \psi (x_2){\bar\psi}(x_3)\sigma^{\mu\nu}\lambda^{0}\psi (x_4)
  |0\rangle \,,
 \label{0.04}\\
 {\cal G}_{T}(x_1,x_2,x_3,x_4) & = &
 \sum_{\alpha = 1}^{N_f^2-1}
 \langle 0|{\bf T}{\bar\psi}(x_1)\sigma_{\mu\nu}\lambda^{\alpha}
 \psi(x_2){\bar\psi}(x_3)\sigma^{\mu\nu}\lambda^{\alpha}\psi (x_4)
 |0\rangle  \,.
 \label{0.05}
\end{eqnarray}
In these expressions, $\lambda^{\alpha}$ is the flavor matrix with
$\alpha = 1,\ldots ,N_{f}^{2}-1$; $\lambda^{0}$ is proportional to the
unit matrix. These four-fermion functions can be related to the axial
charge, and the condensate of these four-fermion functions (with all
the coordinates $x_k$ in the same space-time point) can be considered
as the order parameters of DCSB.

A nonzero condensate of (one of) these four-fermion functions gives
the symmetry breaking pattern
\begin{eqnarray}    \label{symbrpat}
  U_{L}(N_{f})\otimes U_{R}(N_{f}) & \longrightarrow &
    U_{L+R}(N_{f})\otimes Z_{4} \,.
 \label{0.2}
\end{eqnarray}
Here $Z_{4}$ is a nontrivial chiral subgroup which plays a role in
preventing dynamical mass generation for the fermions. It corresponds
to the discrete chiral transformation
\begin{eqnarray}
 \psi(x) &\rightarrow& {\rm e}^{i\frac{k}{2}\pi\gamma_5} \psi(x) \,,
\end{eqnarray}
and therefore the fermions remain massless in such a breaking of the
chiral symmetry.

Note that in principle there are a lot more four-fermion functions
which break the chiral symmetry. In the case of one fermion flavor,
$N_{f}=1$, there are in total 54 independent functions which all lead
to the same chiral symmetry breaking pattern, Eq.~(\ref{symbrpat}),
but the ones above, which we will call $S+P$ and $T$ type four-fermion
functions, are the only ones which preserve parity and Lorentz
symmetry.  The other 52 independent functions are not all
parity-invariant or Lorentz scalars, which means they will not receive
nonzero condensates, assuming that the Lorentz symmetry and parity
remains unbroken. However, it is not excluded that (some of) the
nonlocal four-fermion functions become nonzero, in which case they
might couple to the $S+P$ and $T$ type functions. We leave this
problem to future investigations and for simplicity we only consider
the nonperturbative four-point functions which are flavor-, Lorentz-
and parity-invariant.

This scenario can easily be generalized to condensates of $2n$
fermions, breaking the chiral symmetry into
\begin{eqnarray}
  U_{L}(N_{f})\otimes U_{R}(N_{f}) & \longrightarrow &
    U_{L+R}(N_{f})\otimes Z_{2n} \,,
\end{eqnarray}
with the nontrivial chiral subgroup $Z_{2n}$, corresponding
to the discrete chiral transformation
\begin{eqnarray}
  \psi(x) &\rightarrow&
  {\rm e}^{i\frac{k}{n}\pi\gamma_5} \psi(x) \,,
\end{eqnarray}
unbroken.

%%%%%%%%%%%%%%%%%%%%%%%%%%%%%%%%%%%%%%%%%%%%%%%%%%%%%%%%%%%%%%%%%%%%%%%%%%%%%
\subsection{Goldstone Bosons}

It is wellknown that if a continuous symmetry of the Lagrangian is
broken dynamically, there will appear some massless bosons, the
socalled Goldstone or Nambu--Goldstone bosons \cite{Nambu,Goldstone}.
These Goldstone bosons couple both to the conserved current which is
associated with the symmetry, and to some fields $\phi$ which
characterize the breaking of the symmetry. Note that these fields
$\phi$, which can be regarded as the order parameters of the phase
transition, can be either composite or elementary.

In the case of dynamical breaking of chiral symmetry, the Goldstone
bosons are related to the chiral-symmetry breaking $n$-point functions
through the axial (or chiral) Ward--Takahashi identity. The conserved
chiral current is
\begin{eqnarray}
  j^\mu_5(x) &=& \bar\psi(x) \gamma_5 \gamma^\mu \psi(x) \,,
\end{eqnarray}
and we consider the following Green's function in momentum space
\begin{eqnarray}
  G^\mu_5(q) &=& \int d^4x\,{\rm e}^{iqx}
          \langle 0| j^\mu_5(0) \phi(x) |0\rangle \,,
\end{eqnarray}
where $\phi$ is a (composite) field associated with the chiral order
parameter we are considering\footnote{Here we only write down the
space-time coordinate $x$ which is related to Goldstone pole in
momentum space; in general, $\phi$ can also depend on other space-time
coordinates, which do not relate to the Goldstone pole.}. The breaking
of chiral symmetry means that this order parameter becomes nonzero,
and due to the chiral Ward--Takahashi identity this leads to a pole at
zero momentum in the above Green's function; in the limit $q_\mu
\rightarrow 0$ this Green's function behaves as
\begin{eqnarray}
  G^\mu_5(q) &\sim & \frac{q^\mu}{q^2} \int dx\,{\rm e}^{iqx}
       i\partial_\nu \langle 0| j^\nu_5(0) \phi(x) |0\rangle  \\
     &\sim & \frac{-q^\mu}{q^2} \langle 0| \delta\phi |0\rangle \,.
\end{eqnarray}

To be more specific, we first consider conventional chiral symmetry
breaking, with a dynamically generated mass for the fermions. In that
case, the composite field $\phi(x-y)$ is $\psi(x)\bar\psi(y)$, and the
chiral Ward--Takahashi identity leads to
\begin{eqnarray}
 (p-k)_\mu G^\mu_5(p-k) &=& S(p) \gamma_5 + \gamma_5 S(k)  \,,
\end{eqnarray}
where $S(p)$ is the full fermion propagator. From this equation, we
see immediately that a dynamically generated mass for the fermions
will lead to a massless pole in the pseudoscalar Green's function
$G^\mu_5(p-k)$ if $p \rightarrow k$; this pole corresponds to the
massless Goldstone boson.

In the case of dynamical chiral symmetry breaking through the
four-point functions, as we are considering here, the situation is
somewhat more complicated since we have to consider a composite field
$\phi$ which consists of four fermion fields. Using the chiral
Ward--Takahashi identity, one can show that the Green's functions
$G^5_\mu(q,p_i)$ have to satisfy the relation
\begin{eqnarray}                 \label{axward}
 q_\mu (G^\mu_5(q,p_i))_{abcd} &=&
    (\gamma_5)_{aa'} {\cal G}_{a'bcd}(p_1+q,p_2,p_3,p_4)
  + {\cal G}_{ab'cd}(p_1,p_2+q,p_3,p_4) (\gamma_5)_{b'b}
\nonumber \\ &&
  + (\gamma_5)_{cc'} {\cal G}_{abc'd}(p_1,p_2,p_3+q,p_4)
  + {\cal G}_{abcd'}(p_1,p_2,p_3,p_4+q) (\gamma_5)_{dd'} \,,
\end{eqnarray}
where ${\cal G}_{abcd}(p_1,p_2,p_3,p_4)$ is the Fourier transform of a
four-fermion function, with momentum conservation $\sum p_i = 0$. In
the limit $q_\mu \rightarrow 0$, this leads to massless poles, if
either the $S+P$ type or the $T$ type four-fermion function
\begin{eqnarray}
 {\cal G}_{S+P}(p_1,p_2,p_3,p_4) & = &
 \int{d^{4}x_{1}\cdots d^{4}x_{4}} \,
   {\rm e}^{-i\sum{p_{i}x_{i}}}
 \langle 0|{\bf T}{\bar\psi}(x_1)\psi (x_2){\bar\psi}(x_3)\psi (x_4)
 |0\rangle  \nonumber\\&& +
 \int{d^{4}x_{1} \cdots d^{4}x_{4}} \,
   {\rm e}^{-i\sum{p_{i}x_{i}}}
   \langle 0|{\bf T}{\bar\psi}(x_1)\gamma_{5}\psi (x_2)
   {\bar\psi}(x_3)\gamma_{5}\psi (x_4)|0\rangle \,,
\\
 {\cal G}_{T}(p_1,p_2,p_3,p_4) & = &
 \int{d^{4}x_{1} \cdots d^{4}x_{4}} \,
   {\rm e}^{-i\sum{p_{i}x_{i}}}
 \langle 0|{\bf T}{\bar\psi}(x_1)\sigma_{\mu\nu}\psi (x_2)
  {\bar\psi}(x_3)\sigma^{\mu\nu}\psi (x_4)|0\rangle  \,,
\end{eqnarray}
is nonzero. This can be seen easily by decomposing the Green's
function $G_5^\mu$ into
\begin{eqnarray}
 G^\mu_5(q,p_i) &=& \frac{q_\mu}{q^2} \Big(\chi_{S+P}(q,p_i)
       \left( I \otimes \gamma_5 + \gamma_5 \otimes I \right)
      + \chi_{T}(q,p_i) \,(\gamma_5 \sigma_{\rho\sigma})
         \otimes \sigma^{\rho\sigma} \Big) + \ldots
\end{eqnarray}
so Eq.~(\ref{axward}) leads to
\begin{eqnarray}
 \chi_{S+P}(q,p_i) &=& {\cal G}_{S+P}(p_1+q,p_2,p_3,p_4)
  + {\cal G}_{S+P}(p_1,p_2+q,p_3,p_4)
\nonumber \\ &&
  + {\cal G}_{S+P}(p_1,p_2,p_3+q,p_4)
  + {\cal G}_{S+P}(p_1,p_2,p_3,p_4+q)  \,,
\\
 \chi_{T}(q,p_i) &=& {\cal G}_{T}(p_1+q,p_2,p_3,p_4)
  + {\cal G}_{T}(p_1,p_2+q,p_3,p_4)
\nonumber \\ &&
  + {\cal G}_{T}(p_1,p_2,p_3+q,p_4)
  + {\cal G}_{T}(p_1,p_2,p_3,p_4+q) \,.
\end{eqnarray}
Thus massless Goldstone bosons appear as soon as any of these
nonperturbative four-point functions becomes nonzero.

If we restrict ourselves to {\em local} composite fields $\phi$ we get
a similar expression, but now involving the {\it condensates} of these
four-point functions. Defining the condensate as
\begin{eqnarray}
  \int\frac{d^{4}p_{1}d^{4}p_{2}d^{4}p_{3}d^{4}p_{4}}{(2\pi )^{12}}
   {\cal G}_{i}(p_{1},p_{2},p_{3},p_{4}) \delta(p_1+p_2+p_3+p_4) \,,
\end{eqnarray}
where $i$ denotes either $S+P$ or $T$, the chiral Ward--Takahashi
identity relates this condensate to the Goldstone bosons. In momentum
space this finally leads to massless poles for $q^2 \rightarrow 0$
\begin{eqnarray}
 \chi_{S+P}(0) &=&
 4 \int\frac{d^{4}p_{1}d^{4}p_{2}d^{4}p_{3}}{(2\pi )^{12}}
 {\cal G}_{S+P}(p_1,p_2,p_3,-p_1-p_2-p_3) \,,
\\
 \chi_{T}(0) &=&
 4 \int\frac{d^{4}p_{1}d^{4}p_{2}d^{4}p_{3}}{(2\pi )^{12}}
 {\cal G}_{T}(p_1,p_2,p_3,-p_1-p_2-p_3) \,.
\end{eqnarray}
Thus the condensates of the four-point functions we are considering
here are directly related to the existence of massless Goldstone
bosons.

Note that both a nonzero condensate and a nonzero four-point function
would signal dynamical chiral symmetry breaking, and in principle it
is conceivable that the condensate is zero, and the four-point
function not; the other way around (no nonzero four-point function,
but only a nonzero condensate) is obviously impossible. In this paper,
instead of four-fermion condensate, we directly calculate four-point
function and take it as a generalized order parameter.

The generalization to the case of $N$ fermion flavors is in principle
straightforward; also the generalization to higher $n$-point functions
is a trivial exercise.

%%%%%%%%%%%%%%%%%%%%%%%%%%%%%%%%%%%%%%%%%%%%%%%%%%%%%%%%%%%%%%%%%%%%%%%%%%%%%
\section{Effective Action}
\label{seceffact}
%%%%%%%%%%%%%%%%%%%%%%%%%%%%%%%%%%%%%%%%%%%%%%%%%%%%%%%%%%%%%%%%%%%%%%%%%%%%%

In order to calculate the chirality-changing four-fermion functions,
we need some nonperturbative calculation scheme. For this purpose, we
use a loop expansion of the effective action
\cite{effact,Jac73,CJT74}. For simplicity we will restrict ourselves
to just one fermion flavor, so the generating functional in Minkowski
space is
\begin{eqnarray}
 Z[{\bar I},I,J] \; = \; {\rm e}^{W[{\bar I},I,J]}
 & = & {\int}{\cal D}\bar\psi {\cal D}\psi {\cal D}A_{\mu} \;
       {\rm e}^{i{\int}d^{4}x[ {\cal L}(A)
       + {\bar\psi}(i\partial\!\!{\!}/-eA\!\!{\!}/)\psi
       + {\bar I}{\psi}+{\bar\psi}I + J_{\mu}A^{\mu}]}  \,,
\label{genfun}
\end{eqnarray}
where ${\cal L}(A)$ is the Lagrangian for the photon field
\begin{eqnarray}
 {\cal L}(A) & = & - \textstyle{\frac14} F_{\mu\nu}F^{\mu\nu}
    - \textstyle{\frac{1}{2a}} \left(\partial_\mu A^\mu\right)^2 \,,
\end{eqnarray}
in a general covariant gauge. In order to calculate the effective
action, we start by integrating out the photon field, which gives
\begin{eqnarray}             \label{501}
 \lefteqn{
   \int{\cal D}A_{\mu}{\rm e}^{{i\int}d^{4}x[{\cal L}(A)
    +(-e\bar{\psi}\gamma^{\mu}\psi+J^{\mu})A_{\mu}]} \; =}
\nonumber \\
& = & \exp\bigg[\frac{-i}{2}{\int}d^{4}xd^{4}y
 \left(-e\bar{\psi}(x)\gamma^{\mu}\psi (x)+J^{\mu}(x)\right)
   D_{\mu\nu}(x,y) \left(-e\bar{\psi}(y)\gamma^{\nu}\psi (y)
   +J^{\nu}(y)\right)\bigg] \,,
\end{eqnarray}
where $D_{\mu\nu}(x,y)$ is the bare photon propagator
\begin{eqnarray}              \label{501.1}
 D_{\mu\nu}(x,y) &\equiv&
 \frac{1}{\partial^{2}} \left[ g_{\mu\nu} -
 \frac{\partial_{\mu}\partial_{\nu}}{\partial^{2}}
              \left(1+ a \right)\right] \delta^4(x-y)  \,,
\end{eqnarray}
in which the derivative $\partial_\mu$ is acting on the coordinate
$x$, and $a$ is the gauge-fixing parameter.

Next, we introduce a generalized Green's function containing four
space-time points, with spinor indices $a,b,c$, and $d$
\begin{eqnarray}               \label{502.1}
  \overline{D}^{ab}_{cd}(x,x',y,y') &\equiv&
  D_{\mu\nu}(x,y)\gamma^{\mu}_{ad}\gamma^{\nu}_{bc}
  \delta^4(x'-y)\delta^4(y'-x)  \,,
\end{eqnarray}
such that
\begin{eqnarray}               \label{502}
\lefteqn{ -D_{\mu\nu}(x,y)
  \left(\bar{\psi}(x)\gamma^{\mu}{\psi}(x)\right)_{aa}
  \left(\bar{\psi}(y)\gamma^{\nu}{\psi}(y)\right)_{bb} \; =}
\nonumber\\
 &=&{\int}d^{4}x'd^{4}y'\overline{D}^{ab}_{cd}(x,x',y,y')
 \bar{\psi}^{a}(x){\psi}^{c}(x') \bar{\psi}^{b}(y){\psi}^{d}(y') \,.
\end{eqnarray}
Note that in the case with $N_f$ fermion flavors, we just have to add
additional indices to label the different flavors. Thus the generating
functional can be written as
\begin{eqnarray}               \label{503}
Z[{\bar I},I,J] & = &
  \int{\cal D}\psi{\cal D}\bar{\psi}\exp\,i \Bigg[ {\int}d^{4}x
  \left(\bar{\psi}i\partial\!\!\!/{\psi}+{\bar I}\psi+\bar{\psi}I
  \right) \nonumber\\
&& + \frac{e^2}{2} {\int}d^{4}xd^{4}x'd^{4}yd^{4}y'
   \overline{D}^{ab}_{cd}(x,x',y,y')
  \bar{\psi}^{a}(x){\psi}^{c}(x') \bar{\psi}^{b}(y){\psi}^{d}(y')
 \nonumber\\
&& - \frac{1}{2} {\int}d^{4}xd^{4}y D_{\mu\nu}(x,y)
     \left(-2e\bar{\psi}(x)\gamma^{\mu}\psi (x)J^{\nu}(y)
     + J^{\mu}(x)J^{\nu}(y)\right) \Bigg]  \,.
\end{eqnarray}
Inserting a Gaussian integral of a bilocal auxiliary field
$\chi_{ab}(x,x')$
\begin{eqnarray}               \label{504}
 \hbox{const} &=& \int{\cal D}\chi \exp\,\frac{-i}{2}
  \Bigg[{\int}d^{4}xd^{4}x'd^{4}yd^{4}y'
  \overline{D}^{ab}_{cd}(x,x',y,y')\times
\nonumber\\
&& \left(\chi^{ac}(x,x')-e\bar{\psi}^{a}(x){\psi}^{c}(x')\right)
   \left(\chi^{bd}(y,y')-e\bar{\psi}^{b}(y){\psi}^{d}(y')\right)
  \Bigg] \,,
\end{eqnarray}
gives for the generating functional
\begin{eqnarray}               \label{505}
\lefteqn{ Z[{\bar I},I,J] \; = \;
 \int{\cal D}\chi{\cal D}\psi{\cal D}\bar{\psi}\exp\,i \Bigg[
  {\int}d^{4}x \Big( \bar{\psi}(x)i\partial\!\!\!/_{x} \psi(x)
   + {\bar I}(x)\psi(x) +\bar{\psi}(x)I(x)\Big) }
\nonumber\\ &&
  + {\int}d^{4}xd^{4}x'd^{4}yd^{4}y'
    \overline{D}^{ab}_{cd}(x,x',y,y')
    \left( e \bar{\psi}^{a}(x) \chi^{bd}(y,y') \psi^{c}(x')
        -  \frac{1}{2} \chi^{ac}(x,x')\chi^{bd}(y,y')  \right)
\nonumber\\ &&
  + e{\int}d^{4}x d^{4}y \bar{\psi}(x)
    \gamma^{\mu} D_{\mu\nu}(x,y) J^{\nu}(y) \psi(x)
  - \frac{1}{2}{\int}d^{4}xd^{4}y J^{\mu}(x)
  D_{\mu\nu}(x,y) J^{\nu}(y)  \Bigg] \,.
\end{eqnarray}
We see that the fermion field appears quadratically in the generating
functional, so that they can be easily be integrated out. Thus we
arrive at
\begin{eqnarray}               \label{506}
Z[{\bar I},I,J] & = &
  \int{\cal D}\chi {\rm e}^{iS_{\rm{eff}}[\chi,{\bar I},I,J]}
\nonumber \\
 & = & \int{\cal D}\chi \exp\,i\Bigg[
 \frac{-1}{2}{\int}d^{4}xd^{4}x'd^{4}yd^{4}y'
 \overline{D}^{ab}_{cd}(x,x',y,y')\chi^{ac}(x,x')\chi^{bd}(y,y')
\nonumber\\ &&
  + i{\rm Tr}\,{\rm ln}\,S - {\int}d^{4}xd^{4}y
   \left(\frac{1}{2} J^{\mu}(x) D_{\mu\nu}(x,y)J^{\nu}(y)
    + {\bar I}^{a}(x)S_{ab}(x,y)I^{b}(y)\right) \Bigg]  \,,
\end{eqnarray}
in which $S_{ab}(x,y)$ is defined as
\begin{eqnarray}                \label{506.1}
\lefteqn{ S^{-1}_{ac}(x,x') \;\equiv\;} \nonumber\\
&\equiv&  \left( i\partial\!\!\!/_{x}
    + e \gamma^{\mu}\int d^{4}y D_{\mu\nu}(x,y) J^{\nu}(y)
    \right)_{ac}\delta^{4}(x-x')
    + e{\int}d^{4}yd^{4}y'\overline{D}^{ab}_{cd}(x,x',y,y')
    \chi^{bd}(y,y') \nonumber\\
& = & \left( i\partial\!\!\!/_{x}
      + e \gamma^{\mu}\int d^{4}y D_{\mu\nu}(x,y) J^{\nu}(y)
      \right)_{ac}\delta^{4}(x-x')
      + e D_{\mu\nu}(x,x')\gamma^{\mu}_{ad}\gamma^{\nu}_{bc}
      \chi^{bd}(x',x) \,.
\end{eqnarray}
The effective action for the $\chi$-field, induced by Eq.~(\ref{506})
and with external sources, is then given by
\begin{eqnarray}                 \label{507}
\lefteqn{ S_{\rm{eff}}[\chi,{\bar I},I,J]
  \;= \; \frac{-1}{2} {\int}d^{4}xd^{4}x'
  D_{\mu\nu}(x,x')\gamma^{\mu}_{ad}\gamma^{\nu}_{bc}
  \chi^{ac}(x,x')\chi^{bd}(x',x)}
\nonumber\\ &&
 + i{\rm Tr}\,{\rm ln}\,S - {\int}d^{4}xd^{4}y
   \left(\frac{1}{2}J^{\mu}(x)D_{\mu\nu}(x,y)J^{\nu}(y)
               +{\bar I}^{a}(x)S_{ab}(x,y)I^{b}(y)\right) \,,
\end{eqnarray}
where we have used Eq.~(\ref{502.1}). Given the effective action
$S_{\rm{eff}}[\chi,{\bar I},I,J]$ in Eq.~(\ref{507}), the calculation
of various physical quantities concerns the calculation of the
following generating functional
\begin{eqnarray}                 \label{512}
 \tilde{Z}[{\bar I},I,J,\tilde J]
  &=& {\rm e}^{i\tilde{W}[{\bar I},I,J,\tilde J]}
\nonumber\\
  &=& \int{\cal D}{\chi}\exp\,i\left[S_{\rm{eff}}[\chi,{\bar I},I,J]
    + {\int}d^{4}X  {\;\;}\tilde J^{B}(X)\chi_{B}(X) \right]  \,,
\end{eqnarray}
where $\tilde J^{B}(X)$ is an external source coupled to the
$\chi_{B}(X)$ field, and we use a shorthand notation $X$ and $B$ for
the bilocal variables $(x,x')$ and indices $(ab)$. The generating
functional $Z[{\bar I},I,J]$ in Eq.~(\ref{504}) is related to
$\tilde{Z}[{\bar I},I,J,\tilde J]$ by $Z[{\bar I},I,J]=\tilde{Z}[{\bar
I},I,J,\tilde J=0]$. Now we define the classical field
${\hat{\chi}}_B(X)$ by
\begin{eqnarray}                  \label{513}
 {\hat{\chi}}_{B}(X)  &\equiv&
  \frac{{\delta}\tilde{W}[{\bar I},I,J,\tilde J]}{{\delta}
  \tilde J^{B}(X)} \,,
\end{eqnarray}
and the partial Legendre transformation
\begin{eqnarray}                  \label{514}
 \tilde{\Gamma}[{\bar I},I,J,{\hat{\chi}}]
  &=& W[{\bar I},I,J,\tilde J]
      - {\int}d^{4}X{\;\;}\tilde J^{B}(X) {\hat{\chi}}_{B}(X) \,.
\end{eqnarray}
It is easy to see that
\begin{eqnarray}                   \label{515}
 \frac{\delta\tilde{\Gamma}[{\bar I},I,J,{\hat{\chi}}]}
  {\delta{\hat{\chi}}_{B}(X)} & = & -\tilde J^{B}(X)  \,,
\end{eqnarray}
and hence
\begin{eqnarray}                   \label{516}
 W[{\bar I},I,J] \;=\; \tilde{W}[{\bar I},I,J,0]
  &=& \tilde{\Gamma}[{\bar I},I,J,{\hat{\chi}}]
  \Bigg|_{\frac{\delta\tilde{\Gamma}}{\delta{\hat{\chi}}_{B}}=0} \;.
\end{eqnarray}
Therefore what we need to do is to calculate $\tilde{\Gamma}[{\bar
I},I,J,{\hat{\chi}}]$. When we turn off the source for the bilocal
field, $\tilde J^{B}(X)$, Eq.~(\ref{515}) becomes
\begin{eqnarray}                   \label{517}
 \frac{\delta\tilde{\Gamma}[{\bar I},I,J,{\hat{\chi}}]}
 {\delta{\hat{\chi}}_{B}} &=& 0 \,,
\end{eqnarray}
which reflects the stability of the physical vacuum.  The standard
loop expansion formula for $\tilde{\Gamma}[{\bar I},I,J,{\hat{\chi}}]$
is given in \cite{Jac73}, and reduces in the present case to
\begin{eqnarray}                   \label{518}
 \tilde{\Gamma}[{\bar I},I,J,{\hat\chi}] &=&
  S_{\rm{eff}}[{\hat\chi},{\bar I},I,J]
  + \frac{i}{2}{\rm Tr}\,{\rm ln}\left[
  \frac{\delta^{2}S_{\rm{eff}}[{\hat\chi},{\bar I},I,J]}
  {\delta{\hat\chi}_{B_{1}}(X_{1})\delta{\hat\chi}_{B_{2}}(X_{2})}
   \right] + \tilde{\Gamma}_{2} \,,
\end{eqnarray}
where the first term is the tree level contribution, the second term
is the one-loop contribution, and $\tilde{\Gamma}_{2}$ represent
higher-loop contributions, containing the sum over all higher loop 1PI
vacuum diagrams with the propagator
\begin{eqnarray}
 \frac{\delta^{2}S_{\rm{eff}}[{\hat\chi},{\bar I},I,J]}
   {\delta{\hat{\chi}}_{B_{1}}(X_{1})
    \delta{\hat{\chi}}_{B_{2}}(X_{2})} \,,
\end{eqnarray}
and vertices
\begin{eqnarray}
 \frac{\delta^{n}S_{\rm{eff}}[{\hat\chi},{\bar I},I,J]}
 {\delta{\hat{\chi}}_{B_{1}}(X_{1})\ldots
 \delta{\hat{\chi}}_{B_{n}}(X_{n})}  \hspace{2cm} (n\;{\geq}\;3) \,.
\end{eqnarray}
This expansion, Eq.~(\ref{518}), gives systematic rules to calculate
the effective action, and we base our further discussion on this
expansion. For simplicity, we consider the effective potential up to
tree-level order only in our present paper, but {\em in principle} it
is straightforward to go beyond this leading order calculation. At
tree level, when the external source is switched off, we find
\begin{eqnarray}                    \label{519}
  W[{\bar I},I,J] &=& S_{\rm{eff}}[{\hat\chi},{\bar I},I,J] \,,
\end{eqnarray}
at the stationary point of the effective action
\begin{eqnarray}                    \label{520}
  \frac{\delta S_{\rm{eff}}[{\hat\chi},{\bar I},I,J]}
  {\delta {\hat{\chi}^{ac}(x,y)}} &=& 0  \,.
\end{eqnarray}
Using the definitions Eqs.~(\ref{507}) and (\ref{506.1}), the
stationary condition for the effective action, Eq.~(\ref{520}),
leads to
\begin{eqnarray}                    \label{521}
 {\hat\chi}^{ac}(x,x') &=&  -ieS^{ca}(x',x) -
  e{\int}d^{4}yd^{4}y'S^{cb}(x',y)I^{b}(y){\bar I}^{d}(y')S^{da}(y',x)
  \,.
\end{eqnarray}
Combined with Eq.~(\ref{506.1}), we find
\begin{eqnarray}                    \label{521.1}
 \lefteqn{ S^{-1}_{ac}(x,x')
  \; = \;(i\partial\!\!\!/_{x})_{ac}\delta^{4}(x-x')
   + e \gamma^{\mu}_{ac}\delta^4(x-x')
  \int d^{4}y D_{\mu\nu}(x,y)J^{\nu}(y)}
\nonumber\\
 & & - e^{2}D_{\mu\nu}(x,x')\gamma^{\mu}_{ad}\gamma^{\nu}_{bc}
 \left[iS^{db}(x,x')
  + {\int}d^{4}yd^{4}y'S^{da'}(x,y')I^{a'}(y'){\bar I}^{c'}(y)
  S^{c'b}(y,x') \right]  \,.
\end{eqnarray}
Note from Eq.~(\ref{519}) and Eqs.~(\ref{507}) and (\ref{520}), that
the function $S$ is related to the full fermion propagator (two-point
function)
\begin{eqnarray}                     \label{522a}
 \frac{\delta^{2}W[{\bar I},I,J]}{\delta{\bar I}_a(x)\delta I_b(y)}
 &=&  S_{ab}(x,y) - \int d^{4}z{\bar I}_{c}(z)
 \frac{\delta S_{cb}(z,y)}{\delta{\bar I}_{a}(x)}
 \\                                  \label{522b}
 &=&  S_{ab}(x,y) +
 \int d^{4}z\frac{\delta S_{ac}(x,z)}{\delta I_{b}(y)}I_{c}(z)\,,
\end{eqnarray}
where the functional derivatives ${\delta S}/{\delta{\bar I}}$ and
${\delta S}/{\delta I}$ can be calculated using Eq.~(\ref{521.1}).
Combined with Eq.~(\ref{521}), we find that the $\chi$-field is the
full fermion propagator if we switch off the external sources. On the
vacuum, all external sources should be taken to zero, and therefore
Eq.~(\ref{521.1}) gives
\begin{eqnarray}
 S^{-1}_{ac}(x,x')-(i\partial\!\!\!/_{x})_{ac} \delta^{4}(x-x')
  &=& -ie^{2}D_{\mu\nu}(x,x')\gamma^{\mu}_{ad}S^{db}(x,x')
       \gamma^{\nu}_{bc} \,,
\end{eqnarray}
which is just the quenched ladder approximation for the
Schwinger--Dyson equation for the full fermion propagator. By
differentiating the generating functional twice with respect to the
source $J$ and imposing the stationary condition for the effective
action, we can get the Schwinger--Dyson for the photon field.

For the four-point function, we use Eq.~(\ref{522a}) and (\ref{522b})
to relate the four-point function to derivatives of the function $S$.
After putting the sources equal to zero, and defining
\begin{equation}
 G^{acbd}(x,x',y,y')=
 \frac{\delta S_{ac}(x,x')}{\delta{\bar I}_{b}(y)\delta I_{d}(y')}
 \bigg|_{J=0,I={\bar I}=0}
\end{equation}
we find for the full four-point function
\begin{eqnarray}                    \label{527}
{\cal G}^{acbd}(x,x',y,y') \; &\equiv& \;
 \frac{\delta^{4}W[{\bar I},I,J]}
 {\delta{\bar I}_{a}(x)\delta I_{c}(x')
 \delta{\bar I}_{b}(y)\delta I_{d}(y')}\bigg|_{J=0,I={\bar I}=0}
 \nonumber\\
 &=&G^{acbd}(x,x',y,y')-G^{bcad}(y,x',x,y')\nonumber\\
 &=&G^{acbd}(x,x',y,y')-G^{adbc}(x,y',y,x'),
\end{eqnarray}
We have written down all possible expressions, which we can use to
solve the four-point functions. Note that the four-point function
$\cal G$ is written as the sum of the $s$- and $t$-channel components
$G$.

Using Eq.~(\ref{521.1}) we find for the $s$-channel
\begin{eqnarray}                     \label{528}
\lefteqn{ {\int}d^{4}yd^{4}y'S^{-1}_{ab}(x,y)
  G^{bda'c'}(y,y',x_{1},x_{1}') S^{-1}_{dc}(y',x') \;=}
 \nonumber\\ &&
  \gamma^{\mu}_{ad} \left[iG^{dba'c'}(x,x',x_{1},x_{1}')
   + S_{dc'}(x,x_{1}')S_{a'b}(x_{1},x')\right]
    e^{2}D_{\mu\nu}(x,x') \gamma^{\nu}_{bc} \,.
\end{eqnarray}
This is nothing but the (inhomogeneous) Bethe--Salpeter equation. For
the $S+P$ and $T$ type four-point functions in the massless-fermion
phase (or more general: for all nonperturbative four-point functions),
the second term on the right hand side of Eq.~(\ref{528}) vanishes, so
we arrive at following equation, namely
\begin{eqnarray}
\lefteqn{ G^{abcd}(x_1,x_2,x_3,x_4) \; =}
 \nonumber \\
  && -i e^2 {\int}d^{4}y d^{4}y'
   S_{ad'}(x_1,y) \gamma^{\mu}_{d'a'} G^{a'b'cd}(y,y',x_3,x_4)
   \gamma^{\nu}_{b'c'} S_{c'b}(y',x_2) D_{\mu\nu}(y,y') \,.
\end{eqnarray}
This equation is the $s$-channel equation for the nonperturbative
four-fermion functions in the quenched ladder approximation. For the
$t$-channel equation, we can get a similar equation in exactly the
same way. This $t$-channel equation however is completely equivalent
to the $s$-channel equation, and it can also be obtained by applying a
Fierz transformation to the $s$-channel equation.

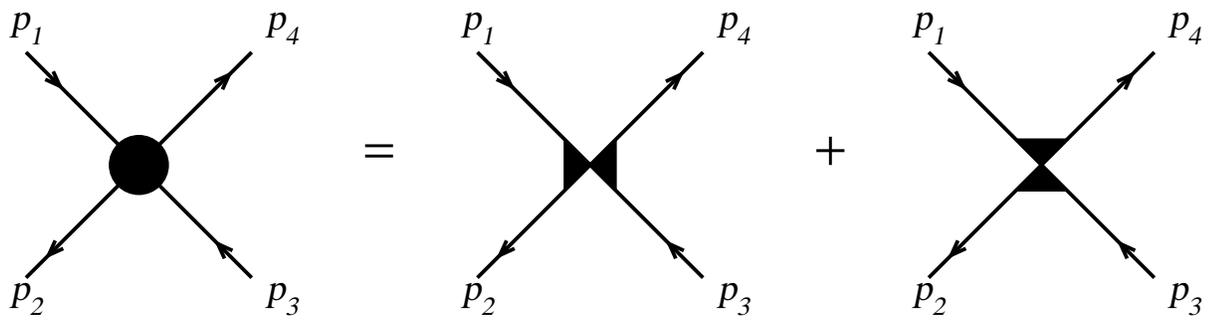
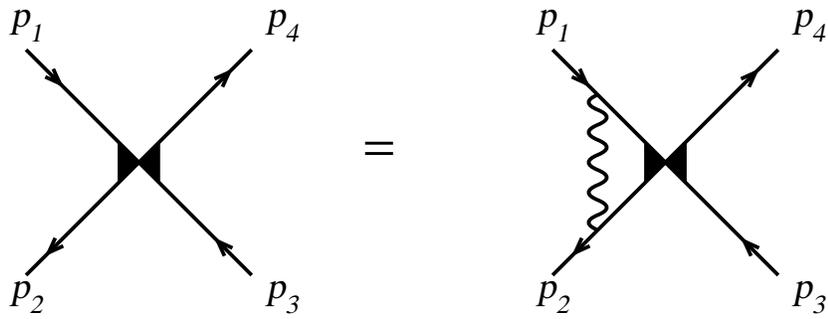
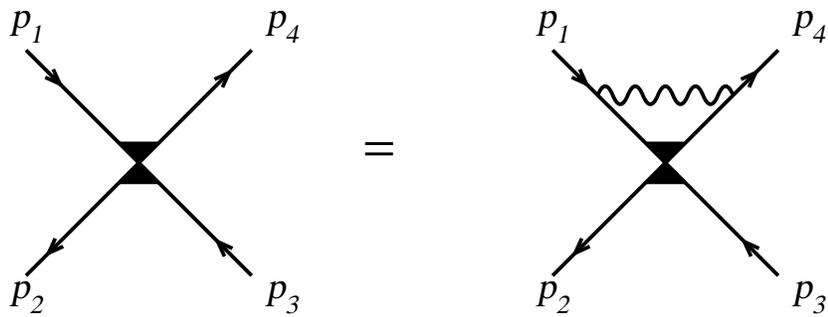
\begin{figure}
 \caption{Our truncated equation for the nonperturbative
        four-fermion function (diagrammatically):
        (a) ${\cal G}(p_1,p_2,p_3,p_4)$ decomposed
        into the $s$- and $t$-channel components $G(p_1,p_2,p_3,p_4)$;
        (b) the equation for the $s$-channel component;
        and (c) the equation for the $t$-channel.}
 \label{figsteqs}
\end{figure}

Next we make a Fourier transformation to momentum space. The equation
for the $s$-channel four-point function becomes
\begin{eqnarray}
 \lefteqn{ G^{abcd}(p_1,p_2,p_3,p_4)  \; =}
 \nonumber \\
  && i e^2 {\int}\frac{d^{4}q}{(2\pi)^4}
   S_{ad'}(p_1) \gamma^{\mu}_{d'a'} G^{a'b'cd}(p_1+q,p_2-q,p_3,p_4)
   \gamma^{\nu}_{b'c'} S_{c'b}(-p_2) D_{\mu\nu}(q) \,,
\end{eqnarray}
where $D_{\mu\nu}(q)$ is the bare photon propagator in momentum
space
\begin{eqnarray}
  D_{\mu\nu}(q) &=&  \frac{-1}{q^2}
  \left[g_{\mu\nu}
  - \left(1 + a\right)\frac{q_\mu q_\nu}{q^2}\right] \,.
\end{eqnarray}
As mentioned before, the $t$-channel equation is completely equivalent
to this $s$-channel equation. For the complete nonperturbative
four-point functions in momentum space, ${\cal
G}(p_{1},p_{2},p_{3},p_{4})$, we thus find
\begin{eqnarray}
 {\cal G}^{abcd}(p_{1},p_{2},p_{3},p_{4})
   &=&
  G^{abcd}(p_1,p_2,p_3,p_4)-G^{cbad}(p_3,p_2,p_1,p_4) \,,
 \\ &=&
  G^{abcd}(p_1,p_2,p_3,p_4)-G^{adcb}(p_1,p_4,p_3,p_2) \,.
\end{eqnarray}
So finally we have the leading order equations for the nonperturbative
four-fermion functions, which is shown in diagrams Fig.~\ref{figsteqs}.
For the perturbative four-fermion functions ones we have to add the
tree contribution coming from the second term in the right hand side
of Eq.(\ref{528}).

%%%%%%%%%%%%%%%%%%%%%%%%%%%%%%%%%%%%%%%%%%%%%%%%%%%%%%%%%%%%%%%%%%%%%%%%%%%
\section{Calculation of the Four-Point Functions}
\label{seccalc}
%%%%%%%%%%%%%%%%%%%%%%%%%%%%%%%%%%%%%%%%%%%%%%%%%%%%%%%%%%%%%%%%%%%%%%%%%%%%%

One can show that the perturbative four-fermion functions do not
contribute at all to the nonperturbative ones, so the nonperturbative
four-point functions form a closed set of equations, at least below
the critical coupling for dynamical mass generation. Furthermore,
since we assume that all the other types of nonperturbative four-point
functions (e.g. the ones which break parity or Lorentz invariance)
remain zero, and we only have to deal with the $S+P$ and $T$ type
four-point functions
\begin{eqnarray}
 {\cal G}_{S+P}(x_1,x_2,x_3,x_4) & = &
  \langle 0|{\bf T}{\bar\psi}(x_1)\psi (x_2)
  {\bar\psi}(x_3)\psi (x_4)|0\rangle
 \nonumber \\
&&  + \langle 0|{\bf T}{\bar\psi}(x_1)\gamma_{5}\psi (x_2)
  {\bar\psi}(x_3)\gamma_{5}\psi (x_4)|0\rangle   \,,
\\
 {\cal G}_{T}(x_1,x_2,x_3,x_4) & = &
  \langle 0|{\bf T}{\bar\psi}(x_1)\sigma_{\mu\nu}\psi (x_2)
  {\bar\psi}(x_3)\sigma^{\mu\nu}\psi (x_4)|0\rangle  \,.
\end{eqnarray}
Using the general nonperturbative equation for the four-fermion
function derived in the previous section, we can project out the
equations for these $S+P$ and $T$ type four-point functions, which
form closed set of linear integral equations.

The four-fermion functions $\cal G_{S+P}$ and $\cal G_{T}$ are related
to the functions $G_{S+P}$ and $G_{T}$ by
\begin{eqnarray}
{\cal G}_{S+P}(p_{1},p_{2},p_{3},p_{4})
 &=&G_{S+P}(p_{1},p_{2},p_{3},p_{4})
 -\frac{1}{2}G_{S+P}(p_{3},p_{2},p_{1},p_{4})
 -3G_{T}(p_{3},p_{2},p_{1},p_{4})\\
 \label{sym1}
 &=&G_{S+P}(p_{1},p_{2},p_{3},p_{4})
 -\frac{1}{2}G_{S+P}(p_{1},p_{4},p_{3},p_{2})
 -3G_{T}(p_{1},p_{4},p_{3},p_{2}) \,,
 \label{sym2}\\
 {\cal G}_{T}(p_{1},p_{2},p_{3},p_{4})
 &=&G_{T}(p_{1},p_{2},p_{3},p_{4})
 -\frac{1}{4}G_{S+P}(p_{3},p_{2},p_{1},p_{4})+
 \frac{1}{4}G_{T}(p_{3},p_{2},p_{1},p_{4})
 \label{sym3}\\
 &=&G_{T}(p_{1},p_{2},p_{3},p_{4})
 -\frac{1}{4}G_{S+P}(p_{1},p_{4},p_{3},p_{2})+
 \frac{1}{4}G_{T}(p_{1},p_{4},p_{3},p_{2})  \,.
 \label{sym4}
\end{eqnarray}
After the projection and reduction of the $\gamma$-algebra, we get for
the $s$-channel $S+P$ and $T$ type four-fermion functions
\begin{eqnarray}
 G_{S+P}(p_{1},p_{2},p_{3},p_{4}) &=&
 ie^{2}\frac{p_{1}\cdot p_{2}}{p_{1}^{2}p_{2}^{2}}\int\frac{d^{4}q}
   {(2\pi )^{4}}D^{\mu}_{\mu}(q)G_{S+P}(p_{1}-q,p_{2}+q,p_{3},p_{4})
  \,, \label{sps}\\
 G_{T}(p_{1},p_{2},p_{3},p_{4})
 &=& ie^{2}\frac{4p_{1}^{\mu}p_{2}^{\nu}-p_{1}\cdot p_{2}g^{\mu\nu}}
   {3p_{1}^{2}p_{2}^{2}}\int\frac{d^{4}q}{(2\pi )^{4}}D_{\mu\nu}(q)
   G_{T}(p_{1}-q,p_{2}+q,p_{3},p_{4}) \,.
 \label{ts}
\end{eqnarray}
where $D_{\mu}^{\mu}(q)$ is the bare photon propagator. Remember that
in $G_{S+P}(p_1,p_2,p_3,p_4)$ and $G_{T}(p_1,p_2,p_3,p_4)$, momentum
conservation condition $p_1+p_2+p_3+p_4=0$ should be satisfied. Since
the $t$-channel equations follow from these $s$-channel equations, it
is enough to analyze these two $s$-channel equations, at least for the
question whether or not there is a critical coupling.

We also have to choose gauge. If we take the Feynman gauge, then the
equation Eq.~(\ref{ts}) gives immediately
$G_{T}(p_{1},p_{2},p_{3},p_{4})=0$, whereas the equation for
$G_{S+P}$, Eq.~(\ref{sps}), leads automatically to
$G_{S+P}(p_{1},p_{2},p_{3},p_{4})=0$ in the Yennie gauge. These
results are obviously artifacts of the specific gauges, and one should
choose the ``correct'' gauge in order to get reasonable results. It is
known that the quenched ladder approximation for the fermion
Schwinger--Dyson equation, which also follows from our effective
action, is most reliable in the Landau gauge, $a=0$. Therefore, we
analyze the equations for these four-fermion functions in the Landau
gauge, and use the photon propagator
\begin{eqnarray}
  D_{\mu\nu}(q) &=&  \frac{-1}{q^2}
  \left(g_{\mu\nu} - \frac{q_\mu q_\nu}{q^2}\right)  \,.
\end{eqnarray}

%%%%%%%%%%%%%%%%%%%%%%%%%%%%%%%%%%%%%%%%%%%%%%%%%%%%%%%%%%
\subsection{$G_{S+P}$}

We start by looking at the equation for $G_{S+P}$,
Eq.~(\ref{sps}), which becomes in Euclidean space, after a Wick
rotation,
\begin{eqnarray}           %{spseucl}
G_{S+P}(p_{1},p_{2},p_{3},p_{4}) &=&
  -3e^{2}\frac{p_{1}\cdot p_{2}}
    {p_{1}^{2}p_{2}^{2}}\int\frac{d^{4}q}{(2\pi )^{4}}
    \frac{1}{q^2} G_{S+P}(p_{1}-q,p_{2}+q,p_{3},p_{4}) \,.
 \label{spseucl}
\end{eqnarray}
We redefine
\begin{eqnarray}                                  %(305.1)
\check{G}_{S+P}(p_{1},p_{2},p_{3},p_{4}) &\equiv&
 G_{S+P}(p_{1},p_{2}-p_{1},p_{3},p_{4})  \,,
 \label{305.1}
\end{eqnarray}
which allows us to rewrite this equation into
\begin{equation}                                     %(306.1)
\check{G}_{S+P}(p_{1},p_{2},p_{3},p_{4})=
 -3e^{2} \frac{p_{1}\cdot (p_{2}-p_{1})}{p_{1}^{2}(p_{2}-p_{1})^{2}}
  \int\frac{d^{4}q}{(2\pi )^{4}}
   \frac{1}{(q-p_{1})^{2}}
   \check{G}_{S+P}(q,p_{2},p_{3},p_{4})  \,.
\label{306.1}
\end{equation}
In order to solve this equation, we first expand ${\hat
G}_{S+P}(p_{1},p_{2},p_{3},p_{4})$ in terms of ${\hat p}_{2}^{\mu}$
\begin{eqnarray}                                     %(307.1)
\check{G}_{S+P}(p_{1},p_{2},p_{3},p_{4}) &=& \sum_{n=0}^{\infty}
 ({\hat p}_{2})_{\mu_{1}}\ldots ({\hat p}_{2})_{\mu_{n}}
 {\tilde G}_{n}^{\mu_{1}\ldots \mu_{n}}(p_{1},p_{2r},p_{3},p_{4}) \,,
\label{307.1}
\end{eqnarray}
where we use the notation $\hat p_i^\mu \equiv p_i^\mu /\sqrt{p_i^2}$
for the angular dependence, and $p_{i\,r}\equiv \sqrt{p_i^2}$
for the absolute value of the momenta. Note also that
${\tilde G}_{n}^{\mu_{1}\ldots \mu_{n}}(p_{1},p_{2r},p_{3},p_{4})$
is symmetric in the indices $\mu_{1},\ldots,\mu_{n}$. Inserting this
in Eq.~(\ref{306.1}) shows immediately that the coefficients
${\tilde G}_{n}^{\mu_{1}\ldots \mu_{n}}(p_{1},p_{2r},p_{3},p_{4})$
have to satisfy
\begin{eqnarray}                                  %(308.1)%(309.1)
 p_{1}^{2}{\tilde G}_{0}(p_{1},p_{2r},p_{3},p_{4}) \;=\;
 3e^{2}\int\frac{d^{4}q}{(2\pi )^{4}}\frac{1}
  {(q-p_{1})^{2}}{\tilde G}_{0}(q,p_{2r},p_{3},p_{4}) \,,
\label{308.1}\\
 -2 p_{1}^{\mu_{1}}p_{1}^{2}{\tilde G}_{0}(p_{1},p_{2r},p_{3},p_{4})
 +p_{1}^{4}{\tilde G}_{1}^{\mu_{1}}(p_{1},p_{2r},p_{3},p_{4})
  \;=\; 3e^{2}\int\frac{d^{4}q}{(2\pi )^{4}}
 \frac{1}{(q-p_{1})^{2}} \times
\nonumber\\
 \Big[-p_{1}^{\mu_{1}}{\tilde G}_{0}(q,p_{2r},p_{3},p_{4})
  +p_{1}^{2}{\tilde G}_{1}^{\mu_{1}}(q,p_{2r},p_{3},p_{4})\Big] \,,
\label{309.1}
\end{eqnarray}
and
\begin{eqnarray}                                   %(310.1)
\lefteqn{\bigg[\delta^{\mu_{1}\mu_{2}}p_{1}^{2}
 {\tilde G}_{n-2}^{\mu_{3}\ldots \mu_{n}}(p_{1},p_{2r},p_{3},p_{4})
 - 2p_{1}^{\mu_{1}}p_{1}^{2}
 {\tilde G}_{n-1}^{\mu_{2}\ldots \mu_{n}}(p_{1},p_{2r},p_{3},p_{4})
 + p_{1}^{4}{\tilde G}_{n}^{\mu_{1}\ldots \mu_{n}}
  (p_{1},p_{2r},p_{3},p_{4}) } \nonumber \\
&&
 + \hbox{ all symmetric permutations in }(\mu_{1}\ldots\mu_{n})\bigg]
\nonumber\\
& = & 3e^{2}\int\frac{d^{4}q}{(2\pi )^{4}}\frac{1}{(q-p_{1})^{2}}
\bigg[-p_{1}^{\mu_{1}}{\tilde G}_{n-1}^{\mu_{2}\ldots \mu_{n}}
(q,p_{2r},p_{3},p_{4})
 +p_{1}^{2}
     {\tilde G}_{n}^{\mu_{1}\ldots \mu_{n}}(q,p_{2r},p_{3},p_{4})
\nonumber\\ &&
+ \hbox{ all symmetric permutations in }(\mu_{1}\ldots\mu_{n})
\bigg] \,,
\label{310.1}
\end{eqnarray}
with $n=2,3,4,\ldots$. The homogeneous part of above equation
satisfies
\begin{eqnarray}                                  %(311.1)
 p_{1}^{2}{\tilde G}_{n}^{\mu_{1}\ldots \mu_{n}}
 (p_{1},p_{2r},p_{3},p_{4})  &=& 3 e^{2}
  \int\frac{d^{4}q}{(2\pi )^{4}} \frac{1}{(q-p_{1})^{2}}
  {\tilde G}_{n}^{\mu_{1}\ldots \mu_{n}}(q,p_{2r},p_{3},p_{4}) \,.
\label{311.1}
\end{eqnarray}
In order to prove that the only solution of the equations
Eqs.~(\ref{308.1})--(\ref{310.1}) is the trivial solution, it is
enough to prove that the only solution of the homogeneous equation
Eq.~(\ref{311.1}) is the trivial solution ${\tilde
G}_{n}^{\mu_{1}\ldots \mu_{n}}(p_{1},p_{2r},p_{3},p_{4})=0$, for all
$n=0,1,2,\ldots$. Note that only a solution which is symmetric in the
Lorentz indices $\mu_1\ldots \mu_n$ will contribute to the four-point
function $\check G_{S+P}$ ; from Eq.~(\ref{307.1}) it follows
immediately that any nonsymmetric part does not contribute.

The angular integration in Eq.~(\ref{311.1}) can be performed
analytically, using an expansion in a kind of generalized Chebyshev
polynomials
\begin{eqnarray}                        %(408.1)%(409.1)
  {\hat U}_{2n}^{\mu_{1}\ldots\mu_{2n}}({\hat q})
 &=& \sum_{k=0}^{n}
 \frac{(-1)^{k}2^{2n-2k}(2n-k)!}{k!(2n-2k)!(2n)!}
 \Big[{\hat q}^{\mu_{1}}\ldots{\hat q}^{\mu_{2n-2k}}
 \delta^{\mu_{2n-2k+1}\mu_{2n-2k+2}}\ldots \delta^{\mu_{2n-1}\mu_{2n}}
 \nonumber\\
 && + {\hbox{ all symmetric permutations in }}(\mu_{1}\ldots\mu_{2n})
  \Big]\,,
\label{408.1}\\
  {\hat U}_{2n+1}^{\mu_{1}\ldots\mu_{2n+1}}({\hat q})
 &=& \sum_{k=0}^{n}
  \frac{(-1)^{k}2^{2n+1-2k}(2n+1-k)!}{k!(2n+1-2k)!(2n+1)!}
  \Big[{\hat q}^{\mu_{1}}\ldots{\hat q}^{\mu_{2n+1-2k}}
 \delta^{\mu_{2n+2-2k}\mu_{2n+3-2k}}\ldots \delta^{\mu_{2n}\mu_{2n+1}}
\nonumber\\
 &&+{\hbox{ all symmetric permutations in }}(\mu_{1}\ldots\mu_{2n+1})
 \Big] \,,
\label{409.1}
\end{eqnarray}
where $\hat q^\mu = q^\mu /\sqrt{q^2}$. These polynomials satisfy the
identity
\begin{eqnarray}                            %(313.1)
\int d\Omega_{q}\frac{{\hat U}_{n}^{\mu_{1}\ldots\mu_{n}}({\hat q})}
 {(q-p_{1})^{2}}
  & = & {\hat U}_{n}^{\mu_{1}\ldots\mu_{n}}({\hat p}_{1})
 \frac{1}{(n+1)p_{1r}q_r}
   \left[\frac{min(p_{1r},q_r)}{max(p_{1r},q_r)}\right]^{n+1} \,,
\label{313.1}
\end{eqnarray}
which is shown in Appendix \ref{angint}, where we have given more
details on these polynomials. Assuming that our four-point function
is smooth enough (or more precisely, has a Taylor-expansion), we
expand\footnote{This expansion can always be made in a unique way,
and can be thought of as an expansion in ${\hat
p}_{1}^{\mu_{1}}\ldots{\hat p}_{1}^{\mu_{m}}$ first, followed by an
expansion of ${\hat p}_{1}^{\mu_{1}}\ldots{\hat p}_{1}^{\mu_{m}}$ in
terms of ${\hat U}_{l}^{\nu_{1}\ldots\nu_{l}}({\hat p}_{1})$ using
Eqs.~(\ref{417}) and (\ref{418}), see the appendix.} the unknown
functions
${\tilde G}^{\mu_{1}\ldots\mu_{n}}_{n}(p_{1},p_{2r},p_{3},p_{4})$
in terms of these generalized Chebyshev polynomials
\begin{eqnarray}                              %(312.1)
 {\tilde G}^{\mu_{1}\ldots\mu_{n}}_{n}(p_{1},p_{2r},p_{3},p_{4})
 &=&  \sum_{l=0}^{\infty}
 \widetilde{G_{l}^n}^{\mu_{1}\ldots\mu_{n}}_{\nu_{1}\ldots\nu_{l}}
 (p_{1r},p_{2r},p_{3},p_{4})
 {\hat U}_{l}^{\nu_{1}\ldots\nu_{l}}({\hat p}_{1}) \,.
\label{312.1}
\end{eqnarray}
After this decomposition, which allows us to perform the angular
integration, Eq.~(\ref{311.1}) has to hold for each function
${\widetilde G_l^n}$ separately, which leads to an equation for each
coefficient
\begin{eqnarray}                   \label{314.1}
 p_{1}^{2}
 \widetilde{G_l^n}^{\mu_{1}\ldots\mu_{n}}_{\nu_{1}\ldots\nu_{l}}
 (p_{1r},p_{2r},p_{3},p_{4})&=&
 \frac{3e^{2}}{16\pi^{2}}\Bigg\{
 \int^{p_{1}^{2}}_{\epsilon^{2}}dq^{2}
 \widetilde{G_l^n}^{\mu_{1}\ldots\mu_{n}}_{\nu_{1}\ldots\nu_{l}}
 (q_{r},p_{2r},p_{3},p_{4})
\nonumber \\
 && + \int^{\Lambda^{2}}_{p_{1}^{2}}dq^{2}
  \widetilde{G_l^n}^{\mu_{1}\ldots\mu_{n}}_{\nu_{1}\ldots\nu_{l}}
 (q_{r},p_{2r},p_{3},p_{4})\frac{p_{1}^{2}}{q^{2}}
 \Bigg\} \,.
\end{eqnarray}

For convenience we will drop all irrelevant indices and the spectator
momenta in the discussion of this equation, defining $g(p)$ as a
scalar function of the absolute value of $p_1$
\begin{eqnarray}
 \widetilde{G_l^n}^{\mu_{1}\ldots\mu_{n}}_{\nu_{1}\ldots\nu_{l}}
 (p_{1r},p_{2r},p_{3},p_{4})&\equiv&  g(p)  \,,
\end{eqnarray}
so the equation we have to consider becomes
\begin{eqnarray}               \label{314.1conv}
 p^{2} g(p) &=&
 \frac{3e^{2}}{16\pi^{2}}\Bigg\{ \int^{p^{2}}_{\epsilon^{2}}dq^{2}
  g(q)  + \int^{\Lambda^{2}}_{p^{2}}dq^{2}
 \frac{p^2}{q^2} g(q) \Bigg\} \,.
\end{eqnarray}
This integral equation is exactly the same as the equation for the
dynamical mass function in quenched ladder QED, which has been studied
extensively \cite{MasNak74,FomMir76,Miretal83}. It can be transformed
into the following differential equation
\begin{eqnarray}                \label{56.1}
 \frac{d}{dp^{2}}\frac{d}{dp^{2}} p^{2}g(p) &=&
 - \frac{3e^{2}}{16\pi^{2}}\frac{g(p)}{p^{2}}  \,.
\end{eqnarray}
There are two independent solutions for the above differential
equation, say $(p^{2})^{b_i}$ with $b_i$ satisfying
\begin{eqnarray}                \label{57.1}
 b_i &=&
 \frac{1}{2}\left[-1\pm\sqrt{1-\frac{3\alpha}{\pi}}\right] \,.
\end{eqnarray}
{}From this equation, it follows immediately that there is a critical
coupling $\alpha_c = {\pi}/{3}$, below which both $b_i$ are real, and
above which they are complex.

So the general solution is
\begin{eqnarray}                                       %(62.1)
 g(p) &=& \sum_{i=1,2} c_{i} \; p^{2b_{i}} \,,
 \label{62.1}
\end{eqnarray}
in terms of the two independent solutions of the differential
equation, with $c_{i}$ some arbitrary functions, depending on the
spectator momenta $p_{2r}$, $p_3$, and $p_4$ in the original equation
Eq.~(\ref{314.1conv}), which needs to be determined. This general
solution is constrained by the integral equation, and therefore, we
substitute this solution back in the original integral equation, which
gives us
\begin{eqnarray}               \label{63.1}
\sum_{i=1,2} c_{i}
 \Bigg[-\frac{\epsilon^{2b_{i}+2}}{ (b_{i}+1)}
   + p^2 \frac{\Lambda^{2b_{i}}}{b_{i}} \Bigg] \,.
  &=& 0
\end{eqnarray}
This boundary condition can only be satisfied above the critical
coupling, if the exponents $b_i$ are complex; below the critical
coupling, both $b_i$ are real numbers, and it is impossible to satisfy
this boundary condition. Assuming there is a nontrivial solution,
Eq.~(\ref{63.1}) is equivalent to the set of equations
\begin{eqnarray}                                      %(63.1)
 \frac{c_1}{c_2}
   &=& - \frac{b_1+1}{b_2+1} \epsilon^{2(b_2-b_1)}  \,, \\
 \frac{c_1}{c_2}
   &=&  - \frac{b_1}{b_2}\Lambda^{2(b_2-b_1)}  \,.
\end{eqnarray}
Below the critical coupling, this set of equations only has a solution
if $\epsilon > \Lambda$, which means the infrared cutoff bigger than
the ultraviolet cutoff. This is obviously not a physical solution, and
therefore the only solution of Eq.~(\ref{56.1}) is the trivial
solution. That means that the only solution of the integral equation
Eq~(\ref{314.1}) is the trivial solution
\begin{eqnarray}
 \widetilde{G_l^n}^{\mu_{1}\ldots\mu_{n}}_{\nu_{1}\ldots\nu_{l}}
 (p_{1r},p_{2r},p_{3},p_{4}) &=&  0  \,.
\end{eqnarray}
Now going back to the original equation for the $S+P$ type
four-fermion function, using the decomposition Eqs.~(\ref{312.1}) and
(\ref{307.1}), and definition Eq.~(\ref{305.1}), we find that only the
trivial solution
\begin{eqnarray}                                      %(34)
  G_{S+P} (p_{1},p_{2},p_{3},p_{4}) &=&  0  \,,
\end{eqnarray}
is allowed.

Above the critical coupling, $\alpha >{\pi}/{3}$, both $b_i$ have a
nonzero imaginary part, which implies oscillating solutions. Instead
of using the complex exponents $b_i$, we write the general solution of
the differential equation Eq.~(\ref{56.1}) as
\begin{eqnarray}
 g(p) & = & \frac{c_{1}}{p}
 \sin\bigg(\sqrt{\frac{3\alpha}{\pi}-1}\ln\frac{p}{\Lambda}\bigg)
 + \frac{c_{2}}{p}\cos\bigg(\sqrt{\frac{3\alpha}{\pi}-1}
 \ln\frac{p}{\Lambda}\bigg) \,,
\end{eqnarray}
where we have used the ultraviolet cut-off $\Lambda$ to set the mass
scale. Substituting this solution back in original integral equation,
we finally find a condition for $\Lambda/\epsilon$
\begin{eqnarray}
  \sqrt{\frac{3\alpha}{\pi}-1}\ln\bigg(\frac{\Lambda}{\epsilon}\bigg)
 &=&  (2n+\frac{1}{2})\pi +\delta \hspace{2cm} \mbox{$n$ integer}\,,
 \\
  \tan\delta &=&\sqrt{\frac{3\alpha}{\pi}-1}   \;\;.
\end{eqnarray}
Now one can always construct nontrivial solutions, depending on this
ratio $\epsilon/\Lambda$. We assume that $\epsilon$, the infrared
cutoff, is related to the energy scale of the dynamically generated
condensate through nonlinear effects, just as in the linearized
quenched ladder approximation for the fermion Schwinger--Dyson
equation, where the infrared cut-off sets the scale for the infrared
mass function, $\epsilon \simeq m(0)$. Thus we allow the infrared
cut-off to be a function of the ultraviolet cut-off, and in general,
the lowest value of $n$ for which $\epsilon < \Lambda$ will correspond
to the ground state solution; higher values of $n$ correspond to
oscillating solutions, i.e. excited states with higher energy.

So there is a critical coupling $\alpha_c = \pi/3$, above which one
can find a nontrivial solution for the homogeneous equation, which is
also the zeroth-order solution for $G(p_{1},p_{2},p_{3},p_{4})$ if one
expand it in terms of $p_{2}^{\mu}$. The only special point is that
the ratio $\epsilon/\Lambda$ must satisfy some constraint. In the
limit that the coupling goes towards the critical coupling,
$\alpha\downarrow \alpha_c$, this ratio goes to infinity, and we can
remove the ultraviolet and infrared cut-offs.

Note that this critical coupling coincides with the critical coupling
for dynamical mass generation for the fermions in the quenched ladder
approximation. This is no surprise, since we noted already that we
arrived at the same integral equation as that for the mass function,
Eq~(\ref{314.1conv}). Looking back at the original equation for
$G_{S+P}$, Eq.~(\ref{spseucl}), we can understand this, because for
$p_1=p_2$ this equation is exactly the same as the linearized quenched
ladder approximation for the fermion propagator.

%%%%%%%%%%%%%%%%%%%%%%%%%%%%%%%%%%%%%%%%%%%%%%%%%%%%%%%%%%
\subsection{$G_{T}$}

Now we calculate $G_T$, Eq.~(\ref{ts}), which becomes in Euclidean
space
\begin{eqnarray}
G_{T}(p_{1},p_{2},p_{3},p_{4}) &=&
  e^{2}\frac{4p_{1}^{\mu}p_{2}^{\nu}-p_{1}\cdot p_{2}\delta^{\mu\nu}}
    {3p_{1}^{2}p_{2}^{2}}\int\frac{d^{4}q}{(2\pi )^{4}}
    \frac{q_\mu q_\nu}{q^4} G_{T}(p_{1}-q,p_{2}+q,p_{3},p_{4}) \,.
 \label{tseucl}
\end{eqnarray}
We can solve this equation in the same way as the $S+P$ type equation,
redefining
\begin{eqnarray}                                  %(305)
\check{G}_{T}(p_{1},p_{2},p_{3},p_{4}) &\equiv&
 G_{T}(p_{1},p_{2}-p_{1},p_{3},p_{4})  \,,
\label{305}
\end{eqnarray}
which allows us to rewrite this equation into
\begin{eqnarray}                                     %(306)
\check{G}_{T}(p_{1},p_{2},p_{3},p_{4})&=&
 e^{2} \frac{4p_{1}^{\mu}(p_{2}-p_{1})^\nu-p_{1}\cdot (p_{2}-p_{1})
   \delta^{\mu\nu}}{3p_{1}^{2}(p_{2}-p_{1})^{2}}
\times\nonumber\\ &&
  \int\frac{d^{4}q}{(2\pi )^{4}}
   \frac{(q-p_{1})_\mu(q-p_{1})_\nu}{(q-p_{1})^{4}}
   \check{G}_{T}(q,p_{2},p_{3},p_{4})  \,.
\label{306}
\end{eqnarray}
Similarly, we expand ${\hat G}_{T}(p_{1},p_{2},p_{3},p_{4})$ in terms
of ${\hat p}_{2}^{\mu}$
\begin{eqnarray}                                     %(307)
\check{G}_{T}(p_{1},p_{2},p_{3},p_{4}) &=& \sum_{n=0}^{\infty}
  ({\hat p}_{2})_{\mu_{1}}\ldots  ({\hat p}_{2})_{\mu_{n}}
 {\tilde G}_{n}^{\mu_{1}\ldots \mu_{n}}(p_{1},p_{2r},p_{3},p_{4}) \,.
\label{307}
\end{eqnarray}
Inserting this in Eq.~(\ref{306}) leads to a set of equations for the
functions ${\tilde G}_{n}^{\mu_{1}\ldots
\mu_{n}}(p_{1},p_{2r},p_{3},p_{4})$, and in order to show that the
only solution for Eq.~(\ref{306}) is the trivial solution, it is
enough to prove that the only solution of the equation
\begin{eqnarray}                                  %(311)
\lefteqn{
 p_{1}^{4}
 {\tilde G}_{n}^{\mu_{1}\ldots \mu_{n}}(p_{1},p_{2r},p_{3},p_{4})
  \;=\;}
\nonumber\\
&=&  e^{2}(-4p_{1}^{\mu}p_{1}^{\nu}+\delta^{\mu\nu}p_{1}^{2})
  \int\frac{d^{4}q}{(2\pi )^{4}}
  \frac{(q_{\mu}-p_{1\mu})(q_{\nu}-p_{1\nu})}{(q-p_{1})^{4}}
  {\tilde G}_{n}^{\mu_{1}\ldots \mu_{n}}(q,p_{2r},p_{3},p_{4}) \,,
\label{311}
\end{eqnarray}
is the trivial solution ${\tilde G}_{n}^{\mu_{1}\ldots
\mu_{n}}(p_{1},p_{2r},p_{3},p_{4})=0$, for all $n=0,1,2,\ldots$,
just as before.

The generalized Chebyshev polynomials satisfy the identity
\begin{eqnarray}                            %(313)
\lefteqn{ \int d\Omega_{q}\frac{(q_{\mu}-p_{1\mu})(q_{\nu}-p_{1\nu})
  (4p_{1}^{\mu}p_{1}^{\nu}-\delta^{\mu\nu}p_{1}^{2})}{(q-p_{1})^{4}}
  {\hat U}_{n}^{\mu_{1}\ldots\mu_{n}}({\hat q}) \; = } \nonumber\\
&=&\delta_{n0}
 + {\hat U}_{n}^{\mu_{1}\ldots\mu_{n}}({\hat p}_{1})\frac{1}{n+1}
 \left[-q^{2}-(p_{1}^{2}-q^{2})p_{1r}\frac{d}{dp_{1r}}\right]
  \frac{1}{p_{1r}q_r}
   \left[\frac{min(p_{1r},q_r)}{max(p_{1r},q_r)}\right]^{n+1} \,,
\label{313}
\end{eqnarray}
so if we expand the unknown functions ${\tilde
G}^{\mu_{1}\ldots\mu_{n}}_{n}(p_{1},p_{2r},p_{3},p_{4})$ in terms of
these polynomials
\begin{eqnarray}                              %(312)
 {\tilde G}^{\mu_{1}\ldots\mu_{n}}_{n}(p_{1},p_{2r},p_{3},p_{4})
 &=&  \sum_{l=0}^{\infty}
 \widetilde{G_{l}^n}^{\mu_{1}\ldots\mu_{n}}_{\nu_{1}\ldots\nu_{l}}
 (p_{1r},p_{2r},p_{3},p_{4})
 {\hat U}_{l}^{\nu_{1}\ldots\nu_{l}}({\hat p}_{1}) \,,
\label{312}
\end{eqnarray}
we can perform the angular integrals in Eq.~(\ref{311}) analytically,
see Appendix \ref{angint}. This gives an equation for each of the
coefficients separately
\begin{eqnarray}                                      %(34)
\lefteqn{
 p_{1}^{4}
 \widetilde{G_l^n}^{\mu_{1}\ldots\mu_{n}}_{\nu_{1}\ldots\nu_{l}}
 (p_{1r},p_{2r},p_{3},p_{4}) \; = \;}
\nonumber \\ &=&
 - \frac{e^{2}}{48\pi^{2}}\Bigg\{
 \int^{p_{1}^{2}}_{\epsilon^{2}}dq^{2}
 \widetilde{G_l^n}^{\mu_{1}\ldots\mu_{n}}_{\nu_{1}\ldots\nu_{l}}
 (q_{r},p_{2r},p_{3},p_{4}) q^{2}
\left[\delta_{l0} +
 \left(\frac{l+2}{l+1}-\frac{l+3}{l+1}\frac{q^{2}}{p_{1}^{2}}\right)
   \left(\frac{q_{r}}{p_{1r}}\right)^{l} \right] \nonumber\\
&& + \int^{\Lambda^{2}}_{p_{1}^{2}}dq^{2}
  \widetilde{G_l^n}^{\mu_{1}\ldots\mu_{n}}_{\nu_{1}\ldots\nu_{l}}
 (q_{r},p_{2r},p_{3},p_{4})q^{2}
\left[\delta_{l0} + \left(-\frac{l}{l+1}\frac{p_{1}^{2}}{q^{2}}
 + \frac{l-1}{l+1}\right)\left(\frac{p_{1r}}{q_{r}}\right)^{l}
 \right]  \Bigg\} \,.
\label{314}
\end{eqnarray}
Again, for convenience we will drop all irrelevant indices and the
spectator momenta in the discussion of this equation, defining $g(p)$
as a scalar function of the absolute value of $p_1$
\begin{eqnarray}                                      %(34)
 \widetilde{G_l^n}^{\mu_{1}\ldots\mu_{n}}_{\nu_{1}\ldots\nu_{l}}
 (p_{1r},p_{2r},p_{3},p_{4})&\equiv&  g(p_1) \,,
\end{eqnarray}
so the equation we have to consider becomes
\begin{eqnarray}                                      %(34)
 p^{4} g(p) &=&
 - \frac{e^{2}}{48\pi^{2}}\Bigg\{ \int^{p^{2}}_{\epsilon^{2}}dq^{2}
 q^2 g(q) \left[\delta_{l0}
 + \left(\frac{l+2}{l+1} - \frac{l+3}{l+1}\frac{q^{2}}{p^{2}}\right)
   \left(\frac{q}{p}\right)^{l} \right] \nonumber\\
&& + \int^{\Lambda^{2}}_{p^{2}}dq^{2}
 q^2 g(q) \left[\delta_{l0} + \left(-\frac{l}{l+1}\frac{p^{2}}{q^{2}}
 + \frac{l-1}{l+1}\right)\left(\frac{p}{q}\right)^{l}\right]
 \Bigg\} \,.
\label{314conv}
\end{eqnarray}
Since the $l=0$ and $l\neq 0$ cases are different, we discuss them
separately.

For $l=0$,the integral equation can be transformed into the following
differential equation
\begin{eqnarray}                                      %(56)
 \frac{d}{dp^{2}}\frac{d}{dp^{2}}p^{6}  g(p) &=&
 -\frac{e^{2}}{16\pi^{2}}p^{2} g(p) \,.
\label{56}
\end{eqnarray}
There are two independent solutions for the above differential
equation, say $(p^{2})^{b-2}$ with $b$ satisfying
\begin{eqnarray}                                       %(57)
 b &=& \frac{1}{2}\left[-1\pm\sqrt{1-\frac{\alpha}{\pi}}\right] \,.
\label{57}
\end{eqnarray}
This means that the general solution of the differential equation can
be written as
\begin{eqnarray}                                       %(62)
 g(p) &=& \sum_{i=1,2} c_{i} \; p^{2b_{i}-2} \,,
 \label{62}
\end{eqnarray}
with $c_{i}$ some arbitrary functions, depending on the spectator
momenta $p_{2r}$, $p_3$, and $p_4$ in the original equation. We
substitute this solution back in the original integral equation in
order to get the correct boundary conditions, which gives us
\begin{eqnarray}
 \sum_{i=1,2} c_{i}
 \Bigg[\frac{\epsilon^{2b_{i}}}{b_{i}-1}
   -\frac{\epsilon^{2b_{i}+2}}{p^{2}b_{i}} \Bigg]
  &=& 0  \,.
\end{eqnarray}
The critical coupling for the nontrivial solution is $\alpha^0_c=\pi$,
as can be seen from Eq.~(\ref{57}), and just as for the $S+P$ type
equation, below the critical coupling this constraint allows only the
trivial solution
\begin{eqnarray}                                       %(58)
 g(p) &=& 0  \,.
 \label{58}
\end{eqnarray}
Above this critical coupling, were the roots $b_i$ become complex, we
can construct a nontrivial solution similar as before.

For $l\neq 0$, the integral equation can be transformed into the
following differential equation
\begin{eqnarray}                                      %(59)
\lefteqn{ \frac{d}{dp^{2}}\frac{d}{dp^{2}}(p^{2})^{l+3}
 \frac{d}{dp^{2}}\frac{d}{dp^{2}}(p^{2})^{-\frac{l}{2}+2}
  g(p)   \;=\;}
\nonumber \\ &=&
 \frac{e^{2}}{48\pi^{2}} \bigg[
 \frac{d}{dp^{2}}\frac{d}{dp^{2}}(p^{2})^{\frac{l}{2}+3} g(p)
  - 3(l+2)\frac{d}{dp^{2}}(p^{2})^{\frac{l}{2}+2} g(p)
  + l(l+2)(p^{2})^{\frac{l}{2}+1}   g(p)\bigg] \,.
 \label{59}
\end{eqnarray}
Now there are four independent solutions for the above differential
equation, say $(p^{2})^{b'-2-\frac{l}{2}}$ with $b'$ satisfying
the equation
\begin{eqnarray}               \label{60}
  f^{\alpha}_{l}(b') \;\equiv\;
 b'(b'+1)(b'-l)(b'-l-1)+\frac{\alpha}{12\pi}
  \big[3b'(b'-l-1)+l(l+2)\big] &=& 0  \,.
\end{eqnarray}
We can calculate some special values $f^{\alpha}_{l}(b')$, to see
where the roots are located, and find for small values of the coupling
\begin{eqnarray}
 f^{\alpha}_{l}(l+1)&=&\frac{\alpha}{12\pi}l(l+2)>0
\nonumber\\
 f^{\alpha}_{l}(l+\frac{1}{2})&=&
  - \frac{1}{4}\left[l^{2}+2l+\frac{3}{4}
  - \frac{\alpha}{3\pi}\left(l^{2}
  + \frac{l}{2}-\frac{3}{4}\right)\right]<0
\nonumber\\
 f^{\alpha}_{l}(l)&=&\frac{\alpha}{12\pi}l(l-1)\geq 0
\nonumber\\
 f^{\alpha}_{l}(0)&=&\frac{\alpha}{12\pi}l(l+2)>0
\label{61}\\
 f^{\alpha}_{l}(-\frac{1}{2})&=&
  - \frac{1}{4}\left[l^{2}+2l+\frac{3}{4}
  - \frac{\alpha}{3\pi}\left(l^{2}
  + \frac{7l}{2}+\frac{9}{4}\right)\right]<0
\nonumber\\
 f^{\alpha}_{l}(-1)&=&\frac{\alpha}{12\pi}(l+2)(l+3)>0
\nonumber
\end{eqnarray}
{}From this it follows immediately that at small values of the coupling
all four roots of the equation are real: two of them are positive and
the other two are negative. In principle, there might be a critical
coupling $\alpha_c$, where two roots become degenerate; such a
critical coupling satisfies the equations
\begin{eqnarray}
  f^{\alpha}_{l}(b') &=& 0 \,,    \\
  f'^{\alpha}_{l}(b') &=& 0 \,.
\end{eqnarray}
This critical coupling will in general depend on the parameter $l$,
and from Eq.~(\ref{61}) we can see immediately that a lower bound for
these critical coupling is
\begin{eqnarray}
  \alpha_c^1      & \geq & {\textstyle\frac{5}{3}}\pi    \,, \\
  \alpha_c^2      & \geq & {\textstyle\frac{105}{53}}\pi \,, \\
  \alpha_c^3      & \geq & {\textstyle\frac{189}{87}}\pi \,, \\
  \cdots \nonumber
\end{eqnarray}
A more detailed calculations shows that the critical coupling for
$l=1$ is indeed $\alpha_c^1 = \frac{5}{3}\pi$.

Again, the general solution can be written as
\begin{eqnarray}
 g(p) &=& \sum_{i=1,2,3,4} c_{i} \; p^{2b'_{i}-4-l}  \,,
\end{eqnarray}
with $c_{i}$ depending on the spectator momenta $p_{2r}$, $p_3$, and
$p_4$ in the original equation Eq.~(\ref{314}), and constrained by the
integral equation. Therefore, we substitute this solution back in the
integral equation, which gives us
\footnote{The reason to put the factor $l-1$ in Eq.~(\ref{lastconstr})
is to keep the equations finite in the case of $l=1$, in which case
$b'=1$ is a solution. We find in this case that in fact
$\frac{l-1}{b'-1}=\frac{1}{1+ \frac{\alpha}{12\pi}}$.}
\begin{eqnarray}                \label{63}
\sum_{i=1}^4 c_{i}
 \Bigg[\frac{l+2}{l+1}\frac{\epsilon^{2b'_{i}}}{p^{l} b'_{i}}
   +\frac{l+3}{l+1}\frac{\epsilon^{2b'_{i}+2}}{p^{l+2}(b'_{i}+1)}
   -\frac{l}{l+1}\frac{p^{l+2}\Lambda^{2b'_{i}-2l-2}}{b'_{i}-l-1}
   +\frac{l-1}{l+1}\frac{p^{l}\Lambda^{2b'_{i}-2l}}{b'_{i}-l}
  \Bigg]  &=& 0 \,,
\end{eqnarray}
which is equivalent to
\begin{eqnarray}
\sum_{i=1}^4 c_{i}\frac{\epsilon^{2b'_{i}}}{b'_{i}}&=&0 \,,\\
\sum_{i=1}^4 c_{i}\frac{\epsilon^{2b'_{i}+2}}{(b'_{i}+1)} &=&0 \,,\\
\sum_{i=1}^4 c_{i}\frac{\Lambda^{2b'_{i}-2l-2}}{b'_{i}-l-1}&=&0 \,,\\
\sum_{i=1}^4 c_{i}\frac{l-1}{b'_{i}-l}
 \frac{\Lambda^{2b'_{i}-2l}}{l+1} &=& 0 \,.
\label{lastconstr}
\end{eqnarray}
We also have the additional constraint that the infrared cut-off is
smaller than the ultraviolet cut-off. This set of equations has no
nontrivial solution below the critical coupling, with all the roots
$b'_i$ real, but above the critical coupling, some of $b'_{i}$ develop
an imaginary part which means that the solution starts oscillating.
Just as in the case of $G_{S+P}$, these oscillations allow for
nontrivial solutions of the integral equation, with $\epsilon <
\Lambda$.

Note, that the critical coupling for $G_{T}$ depends on $l$, and all
of them are bigger than that for $G_{S+P}$ which is $\frac{\pi}{3}$;
the smallest one, corresponding to $l=0$ is $\alpha_c^0 = \pi$.  This
means that one needs a stronger coupling to form a nontrivial $T$ type
function. If the coupling is in the region $\frac{\pi}{3} < \alpha
<\pi$, only $G_{S+P}$ will have a nontrivial solution, while $G_{T}$
still remains zero; at $\alpha = \pi$ this $G_{T}$ might become
nonzero as well.  However, one should keep in mind that these
$G_{S+P}$and $G_{T}$ type function are not identical to the
four-fermion functions ${\cal G}_{S+P}$ and ${\cal G}_{T}$ itself, see
Eqs.~(\ref{sym1})--(\ref{sym4}). So as soon as either $G_{S+P}$ (or
$G_{T}$) becomes nonzero, both the $S+P$ and the $T$ type
nonperturbative four-point functions ${\cal G}_{S+P}$ and ${\cal
G}_{T}$ will become nonzero.

%%%%%%%%%%%%%%%%%%%%%%%%%%%%%%%%%%%%%%%%%%%%%%%%%%%%%%%%%%%%%%%%%%%%%%%%%%%%%
\section{Discussion}
\label{secconc}
%%%%%%%%%%%%%%%%%%%%%%%%%%%%%%%%%%%%%%%%%%%%%%%%%%%%%%%%%%%%%%%%%%%%%%%%%%%%%

We have shown that, in leading order, both the $S+P$ and the $T$ type
four-fermion functions and its condensates are zero below the critical
coupling $\alpha_{c}=\frac{\pi}{3}$ in the massless fermion phase of a
$U(1)$ gauge theory. Above this critical coupling, there can be a
nontrivial solution for these nonperturbative chiral-symmetry breaking
four-point functions. This result is found using an expansion of the
effective action, assuming that the other nonperturbative four-fermion
functions (which are not parity or Lorentz invariant) are zero. It
agrees with the result found by Holdom and Triantaphyllou
\cite{HolTri951}, who used a truncation of the Schwinger--Dyson
equation for the four-fermion functions.

Our result shows that the critical coupling for chiral symmetry
breaking by four-fermion condensates is the same as that for dynamical
mass generation. Therefore, in a strongly coupled abelian gauge
theory, beyond the critical coupling, the chiral symmetry will be
broken through the two-fermion condensate since this is energetically
favored. This means that in this theory, there is no hierarchy of
chiral symmetry breaking, first by the four-fermion condensates, while
the fermions remain massless, and another phase transition, at a lower
energy, where the fermions acquire a mass.

The generalization to the $N_{f}$ flavor case is straightforward, and
does not change our detailed calculations. The only thing we need to
do is to label every $G_i$ function with flavor indices. There are no
flavor mixing effects in the equations, and the flavor symmetry is not
believed to be broken dynamically, so the Green's functions for the
different flavor indices decouple and the form of the equations is
completely same as the ones we discussed. This means that in a
$U_L(N_f) \otimes U_R(N_f)$ gauge theory it seems to be impossible to
use four-fermion functions or its condensates to substitute the
two-fermion condensate as the chiral order parameter.

In a more complicated theory like a chiral gauge theory, as suggested
in \cite{HolKyoto}, there exists in principle the possibility that
chiral symmetry is broken through the forming of four-fermion
condensates while the fermions remain massless. For such a scenario to
be true, there should be some additional symmetry to prevent the
fermions from becoming massive. Whether such a hierarchy of chiral
symmetry breaking can indeed be realized in a chiral gauge theory,
remains an open question; we only showed that in a $U(1)$ gauge theory
the critical coupling for the four-fermion condensates is the same as
that for the two-fermion condensates.

A similar result has been obtained using a truncation of the
Schwinger--Dyson equation for the four-fermion functions
\cite{HolTri951}.  However, the problem with the set of
Schwinger--Dyson equations is that it forms an infinite set of coupled
equations, and it is not at all obvious how to truncate this set of
equations in a consistent way.  Using an approximation which is
similar to the quenched ladder approximation for the fermion
propagator Schwinger--Dyson equation, Holdom and Triantaphyllou have
shown numerically that the critical coupling for chiral symmetry
breaking by four-fermion condensates is almost the same as that for
dynamical mass generation. It is remarkable that both their
calculation and our analytical calculation give the same result (the
same critical coupling), although the set of Feynman diagrams included
in their and our truncation is not the same. This indicates that the
essential diagrams are included in both schemes, and that the exact
details of the truncation are less important. Of course, this should
be confirmed by a study of the next-to-leading order effective
potential, or a more accurate truncation scheme for the
Schwinger--Dyson equation. We should also consider alternative
formulations for the usual effective action, since there are different
ways to introduce the auxiliary fields, which might lead to a
different result.

%%%%%%%%%%%%%%%%%%%%%%%%%%%%%%%%%%%%%%%%%%%%%%%%%%%%%%%%%%%%%%%%%%%%%%%%%%%%%
\section*{Acknowledgement}
%%%%%%%%%%%%%%%%%%%%%%%%%%%%%%%%%%%%%%%%%%%%%%%%%%%%%%%%%%%%%%%%%%%%%%%%%%%%%

We would like to thank Koichi Yamawaki, Yoonbai Kim, and Bob Holdom for
useful and stimulating discussions. This work has been supported in
part by the Nishina Memorial Fund of Japan, the Natural Science Fund
of China, and the Japanese Society for the Promotion of Science.

%%%%%%%%%%%%%%%%%%%%%%%%%%%%%%%%%%%%%%%%%%%%%%%%%%%%%%%%%%%%%%%%%%%%%%%%%%%%%
\appendix
%%%%%%%%%%%%%%%%%%%%%%%%%%%%%%%%%%%%%%%%%%%%%%%%%%%%%%%%%%%%%%%%%%%%%%%%%%%%%
\section{Angular Integration}
\label{angint}
%%%%%%%%%%%%%%%%%%%%%%%%%%%%%%%%%%%%%%%%%%%%%%%%%%%%%%%%%%%%%%%%%%%%%%%%%%%%%

In this appendix we discuss the angular integration. Consider the case
with three independent momenta, and choose the axes (in Euclidean
space) to be
\begin{eqnarray}                \label{201}
 \vec{p}_{1}&=&p_{1}\vec{n}_{1}\nonumber\\
 \vec{p}_{2}&=&p_{2}(\vec{n}_{1}\cos\theta_{p}
 + \vec{n}_{2}\sin\theta_{p})\\
 \vec{q}&=&q(\vec{n}_{1}\cos\theta_{q_{1}}+
 \vec{n}_{2}\sin\theta_{q_{1}}\cos\theta_{q_{2}}+
 \vec{n}_{3}\sin\theta_{q_{1}}\sin\theta_{q_{2}}\cos\phi_{q}+
 \vec{n}_{4}\sin\theta_{q_{1}}\sin\theta_{q_{2}}\sin\phi_{q})\,,
 \nonumber
\end{eqnarray}
where we use the notation $\vec{p}$ for a four-dimensional vector and
$p$ for its absolute value. The angular integration we are interested
in is
\begin{eqnarray}                              %(202)
\lefteqn{\int d\Omega_{q}\frac{(\vec{q}\cdot\vec{p}_{2})^{n}}
{(\vec{q}-\vec{p}_{1})^{2}} \; = \;}
\nonumber\\ &=&
\frac{1}{\pi}\int_{0}^{\pi}d\theta_{q_{1}}\int_{0}^{\pi}
 d\theta_{q_{2}}
 \sin^{2}\theta_{q_{1}}\sin\theta_{q_{2}}\frac{q^{n}p_{2}^{n}
 (\cos\theta_{p}\cos\theta_{q_{1}}
 +\sin\theta_{p}\sin\theta_{q_{1}}\cos\theta_{q_{2}})^{n}}
 {q^{2}+p_{1}^{2}-2qp_{1}\cos\theta_{q_{1}}}
\nonumber\\ &=&
\frac{q^{n}p_{2}^{n}}{2^{n+3}\pi (n+1)\sin\theta_{p}}
 \oint_{unit\;\; circle}
 dt(1-t^{2})t^{-n-2}\frac{(t^{2}e^{-i\theta_{p}}
 +e^{i\theta_{p}})^{n+1}-
(t^{2}e^{i\theta_{p}}+e^{-i\theta_{p}})^{n+1}}
{(q^{2}+p_{1}^{2})t-qp_{1}(1+t^{2})}
\nonumber\\ &=&
 I_{n,0}+I_{n,1} \;\; ,\label{202}
\end{eqnarray}
where the last integration is performed along the unit circle in the
complex $t$ plane. There are two poles inside the circle, one is zero
and the other is $t_{\pm}$, which is defined as
\begin{equation}
t_{\pm}\equiv \frac{p_{1}^{2}+q^{2}\pm
\sqrt{(p_{1}^{2}+q^{2})^{2}-4p_{1}^{2}q^{2}}}{2p_{1}q}
=\Bigg\{\frac{\min[p_{1},q]}{\max[p_{1},q]},
\frac{\max[p_{1},q]}{\min[p_{1},q]}\Bigg\}.
\end{equation}
Their contributions to the integration are denoted by $I_{n,0}$ and
$I_{n,1}$ respectively. We further find
\begin{eqnarray}                                 %(203)
 I_{n,0} &=& -\frac{q^{n-1}p_{2}^{n}}{2^{n+1}(n+1)p_{1}
 \sin\theta_{p}}
 \Bigg\{ \sum_{k=0,n>0}^{[\frac{n-1}{2}]}C_{n+1}^{k}
 \sin[(n+1-2k)\theta_{p}]
 \Bigg[U_{n-1-2k}\left(\frac{p_{1}^{2}+q^{2}}{2p_{1}q}\right)
 \nonumber\\
&&-U_{n+1-2k}\left(\frac{p_{1}^{2}+q^{2}}{2p_{1}q}\right)\Bigg]
 - \sum_{k=[\frac{n}{2}]}^{[\frac{n+1}{2}]} C_{n+1}^{k}
 \sin[(n+1-2k)\theta_{p}] U_{n+1-2k}
 \left(\frac{p_{1}^{2}+q^{2}}{2p_{1}q}\right)
 \Bigg\} \,,
\label{203}\\
 I_{n,1}&=&-\frac{q^{n-1}p_{2}^{n}}{2^{n+1}(n+1)p_{1}\sin\theta_{p}}
 \sum_{k=0}^{n+1} C_{n+1}^{k}t_{-}^{-n-1+2k}
 \sin[(n+1-2k)\theta_{p}] \,,
\end{eqnarray}
where we have used the standard Chebyshev expansion \cite{cheb}
\begin{eqnarray}                                  %(204)
\frac{1}{1-\frac{p_{1}^{2}+q^{2}}{p_{1}q}t+t^{2}}&=&
 \sum_{n=0}^{\infty} t^{n}
 U_{n}\left(\frac{p_{1}^{2}+q^{2}}{2p_{1}q}\right) \,.\label{204}
\end{eqnarray}
Using another property of the Chebyshev functions
\begin{eqnarray}                                   %(205)
\sin [(n+1)\theta]=U_{n}(\cos\theta )\sin\theta \,,\label{205}
\end{eqnarray}
the result of Eq.~(\ref{202}) can be further expressed as
\begin{eqnarray}                                   %(206)%(207)
\int d\Omega_{q}
 \frac{(\vec{q}\cdot\vec{p}_{2})^{2n}}{(q-p_{1})^{2}}
 &=& \frac{q^{2n-1}p_{2}^{2n}}{2^{2n+1}(2n+1)p_{1}}
 \sum_{k=0}^{n} C_{2n+1}^{n-k}U_{2k}(\cos\theta_{p})
  X_{2k}\left(\frac{p_{1}^{2}+q^{2}}{2p_{1}q}\right) \,,
\label{206}\\
 \int d\Omega_{q}
 \frac{(\vec{q}\cdot\vec{p}_{2})^{2n+1}}{(q-p_{1})^{2}}
 &=& \frac{q^{2n}p_{2}^{2n+1}}{2^{2n+2}(2n+2)p_{1}}
 \sum_{k=1}^{n+1} C_{2n+2}^{n+1-k}U_{2k-1}(\cos\theta_{p})
 X_{2k-1}\left(\frac{p_{1}^{2}+q^{2}}{2p_{1}q}\right) \,,
 \label{207}
\end{eqnarray}
where
\begin{eqnarray}
 X_{n}(x) &\equiv& U_{n+1}(x)-U_{n-1}(x)
  + (x-\sqrt{x^{2}-1})^{n+1} - (x+\sqrt{x^{2}-1})^{n+1} \,.
\end{eqnarray}
To calculate $X_{n}(x)$, we consider the quantity
\begin{eqnarray}
 \sum_{n=0}^{\infty}t^{n}X_{n}(x)&=& \sum_{n=0}^{\infty}t^{n}
 \left[U_{n+1}(x)-U_{n-1}(x) + (x-\sqrt{x^{2}-1})^{n+1}
 - (x+\sqrt{x^{2}-1})^{n+1}\right]
 \nonumber\\ &=&
 \frac{1}{t}\left[-1
 +\frac{1-t^{2}-2t\sqrt{x^{2}-1}}{1-2xt+t^{2}}\right]
 \nonumber\\ &=&
 2\frac{x-\sqrt{x^{2}-1}}{1-t(x-\sqrt{x^{2}-1})}
 \nonumber\\ &=&
 2 \sum_{n=0}^{\infty}(x-\sqrt{x^{2}-1})^{n+1}t^{n} \,,
\end{eqnarray}
where Eq.~(\ref{204}) is used. Comparing both sides of the above
equation, we find
\begin{eqnarray}
 X_{n}(x) &=& 2(x-\sqrt{x^{2}-1})^{n+1} \,.
\end{eqnarray}
Now use the definition of the Chebyshev polynomials \cite{cheb}
\begin{eqnarray}                                     %(208)
 U_{n}(x) &=&
 \sum_{k=0}^{[\frac{n}{2}]}(-1)^{k}\frac{(n-k)!}
 {k!(n-2k)!}(2x)^{n-2k} \,, \label{208}
\end{eqnarray}
to obtain
\begin{eqnarray}                                      %(209)%(210)
 \int d\Omega_{q}
 \frac{U_{2n}(\frac{\vec{q}\cdot\vec{p}_{2}}{qp_{2}})}
 {(q-p_{1})^{2}} &=&
 \frac{(-1)^{n}}{p_{1}q} \sum_{k=0}^{n} \alpha_{k}^{(2n)}
 U_{2k}\left(\frac{\vec{p}_{1}\cdot\vec{p}_{2}}{p_{1}p_{2}}\right)
 t_{-}^{2k+1} \,, \label{209}
\\
 \int d\Omega_{q}
 \frac{U_{2n+1}(\frac{\vec{q}\cdot\vec{p}_{2}}{qp_{2}})}
 {(q-p_{1})^{2}}&=&
 \frac{(-1)^{n}}{p_{1}q}\sum_{k=0}^{n} \alpha_{k}^{(2n+1)}
 U_{2k+1}\left(\frac{\vec{p}_{1}\cdot\vec{p}_{2}}{p_{1}p_{2}}\right)
 t_{-}^{2k+2} \,,\label{210}
\end{eqnarray}
in which $\alpha_{k}^{(2n)}$ and $\alpha_{k}^{(2n+1)}$ are defined as
\begin{eqnarray}                                   %(211)%(212)
 \alpha_{k}^{(2n)}&\equiv&
 \sum_{m=k}^{n} (-1)^{m}\frac{(n+m)!}{(n-m)!(m-k)!(m+1+k)!} \,,
\label{211}\\
 \alpha_{k}^{(2n+1)}&\equiv&
 \sum_{m=k}^{n} (-1)^{m}\frac{(n+m+1)!}{(n-m)!(m-k)!(m+2+k)!}\,.
\label{212}
\end{eqnarray}
Finally, to calculate the coefficients $\alpha$, we introduce two
functions
\begin{eqnarray}                                     %(213)%(214)
 F_{n,k}(x) &\equiv&
 \sum_{m=k}^{n} (-1)^{m}\frac{(n+m+2k)!}{(n-m)!m!(m+1+2k)!}x^{m} \,,
\label{213}\\
 {\bar F}_{n,k}(x)&\equiv&
 \sum_{m=k}^{n} (-1)^{m}\frac{(n+m+2k+1)!}{(n-m)!m!(m+2+2k)!}x^{m}\,,
\label{214}
\end{eqnarray}
so we get
\begin{eqnarray}                                      %(215)%(216)
 \alpha_{k}^{(2n)}&=&(-1)^{k}F_{n-k,k}(1) \,,\label{215}\\
 \alpha_{k}^{(2n+1)}&=&(-1)^{k}{\bar F}_{n-k,k}(1) \,.\label{216}
\end{eqnarray}
For $n\leq 1$ we find
\begin{eqnarray}                                          %(217)
 F_{0,k}(x)&=& \frac{1}{2k+1}   \,,\\
 {\bar F}_{0,k}(x)&=&\frac{1}{2k+2}  \,,\\
 F_{1,k}(x)&=& 1-x  \,,\\
 {\bar F}_{1,k}(x)&=&1-x  \,,
\end{eqnarray}
whereas for $n>1$
\begin{eqnarray}                                    %(218)%(219)
 F_{n,k}(x)&=&
 \frac{1}{n!}x^{-2k-1}\frac{d^{n-1}}{dx^{n-1}}[x^{2k+n}(1-x)^{n}] \,,
 \label{218}\\
 {\bar F}_{n,k}(x)&=&
 \frac{1}{n!}x^{-2k-2}\frac{d^{n-1}}{dx^{n-1}}[x^{2k+n+1}(1-x)^{n}]\,,
 \label{219}
\end{eqnarray}
so the total result can be written as
\begin{eqnarray}                                            %(220)
 F_{n,k}(1)&=& \frac{\delta_{n0}}{2k+1}    \,,\\
 {\bar F}_{n,k}(1)&=& \frac{\delta_{n0}}{2k+2}  \,,\\
 \alpha_{k}^{(2n)}&=& \frac{(-1)^{n}}{2n+1} \delta_{nk} \,,\\
 \alpha_{k}^{(2n+1)}&=& \frac{(-1)^{n}}{2n+2}\delta_{nk} \,,
\end{eqnarray}
Substituting this result into Eqs.~(\ref{209}) and (\ref{210}), we
finally get the result
\begin{eqnarray}                             %(405)
\int d\Omega_{q}\frac{U_{n}({\hat q}\cdot{\hat p}_{2})}
 {(\vec{q}-\vec{p}_{1})^{2}}
 &=&  \frac{1}{(n+1)p_{1}q}\Bigg[
      \frac{\min[p_{1},q]}{\max[p_{1},q]}
      \Bigg]^{n+1}U_{n}({\hat p}_{1}\cdot{\hat p}_{2}) \,,
\label{405}
\end{eqnarray}
where ${\hat p}\equiv{\vec{p}}/{p}$ denotes the angular dependence.

Next, we introduce a ``generalized Chebyshev function'' with Lorentz
indices $\mu_1 \ldots \mu_n$ by extracting the ${\hat p}$ dependence
in $U_{n}({\hat q}\cdot{\hat p})$
\begin{eqnarray}                  \label{407}
 U_{n}({\hat q}\cdot{\hat p}) &\equiv&
 {\hat p}_{\mu_{1}}\ldots{\hat p}_{\mu_{n}}
 {\hat U}_{n}^{\mu_{1}\ldots\mu_{n}}({\hat q}) \,,
\end{eqnarray}
which defines the generalized Chebyshev functions
${\hat U}_n^{\mu_1\ldots\mu_n}$ as
\begin{eqnarray}            \label{408}
  {\hat U}_{2n}^{\mu_{1}\ldots\mu_{2n}}({\hat q})
 &=& \sum_{k=0}^{n}
 \frac{(-1)^{k}2^{2n-2k}(2n-k)!}{k!(2n-2k)!(2n)!}
 \Big[{\hat q}^{\mu_{1}}\ldots{\hat q}^{\mu_{2n-2k}}
  \delta^{\mu_{2n-2k+1}\mu_{2n-2k+2}}\ldots
 \delta^{\mu_{2n-1}\mu_{2n}}
 \nonumber\\
 && + {\hbox{ all symmetric permutations in }}(\mu_{1}\ldots\mu_{2n})
 \Big]  \,, \\
  {\hat U}_{2n+1}^{\mu_{1}\ldots\mu_{2n+1}}({\hat q})
 &=& \sum_{k=0}^{n}
  \frac{(-1)^{k}2^{2n+1-2k}(2n+1-k)!}{k!(2n+1-2k)!(2n+1)!}
  \Big[{\hat q}^{\mu_{1}}\ldots{\hat q}^{\mu_{2n+1-2k}}
  \delta^{\mu_{2n+2-2k}\mu_{2n+3-2k}}\ldots \delta^{\mu_{2n}\mu_{2n+1}}
\nonumber\\
 && + {\hbox{ all symmetric permutations in }}(\mu_{1}\ldots\mu_{2n+1})
 \Big] \,.
\end{eqnarray}
Since in Eq.~(\ref{405}), $p_{2}$ is an arbitrary momentum variable,
we can use the expansion Eq.~(\ref{407}) to find
\begin{eqnarray}                  \label{410}
 \int d\Omega_{q}\frac{{\hat U}_{n}^{\mu_{1}\ldots\mu_{n}}({\hat q})}
 {(\vec{q}-\vec{p}_{1})^{2}} &=&
 \frac{1}{(n+1)p_{1}q}\Bigg[ \frac{\min[p_{1},q]}{\max[p_{1},q]}
 \Bigg]^{n+1} {\hat U}_{n}^{\mu_{1}\ldots\mu_{n}}({\hat p}_{1}) \,.
\end{eqnarray}

This formula allows us to perform all of the angular integration which
are needed, rest us to show that we can indeed expand our four-point
functions in terms of these generalized Chebyshev functions. The
Chebyshev polynomials $U_n(x)$ form a complete set of independent
functions, so from Eq.~(\ref{407}), it follows immediately that our
generalized Chebyshev functions are also a complete set for functions
which are totally symmetric in their Lorentz indices. The explicit
expression of this statement can be seen from following identities,
noting that
\begin{eqnarray}                                  %(415)%(416)
 ({\hat p}\cdot{\hat q})^{2n}&=&
 \sum_{k=0}^{n} c_{n,k}U_{2k}({\hat p}\cdot{\hat q})\nonumber\\
 &=& p_{\mu_{1}}\ldots p_{\mu_{2n}}\sum_{k=0}^{n}
 c_{n,k}{\hat U}_{2k}^{\mu_{1}\ldots\mu_{2k}}({\hat q})
 \delta^{\mu_{2k+1}\mu_{2k+2}}\ldots \delta^{\mu_{2n-1}\mu_{2n}}
 \,,\label{415}\\
 ({\hat p}\cdot{\hat q})^{2n+1}&=&
 \sum_{k=0}^{n} c'_{n,k}U_{2k+1}({\hat p}\cdot{\hat q})\nonumber\\
 &=& p_{\mu_{1}}\ldots p_{\mu_{2n+1}}\sum_{k=0}^{n}
 c'_{n,k}{\hat U}_{2k+1}^{\mu_{1}\ldots\mu_{2k+1}}({\hat q})
 \delta^{\mu_{2k+2}\mu_{2k+3}}\ldots \delta^{\mu_{2n}\mu_{2n+1}}
 \,,\label{416}
\end{eqnarray}
which follows directly from Eq.~(\ref{407}) and the fact the Chebyshev
polynomials form a complete set. Now due to the fact that ${\hat p}$
is arbitrary, the Eqs.~(\ref{415}) and (\ref{416}) are equivalent to
\begin{eqnarray}                                  %(417)%(418)
{\hat q}^{\mu_{1}}\ldots{\hat q}^{\mu_{2n}}&=&
 \sum_{k=0}^{n} \frac{c_{n,k}}{(2n)!}
 \Big[{\hat U}_{2k}^{\mu_{1}\ldots\mu_{2k}}({\hat q})
 \delta^{\mu_{2k+1}\mu_{2k+2}}\ldots \delta^{\mu_{2n-1}\mu_{2n}}
 \nonumber\\ &&
 + {\hbox{ all symmetric permutations in }}(\mu_{1}\ldots\mu_{2n})\Big]
 \label{417}\\
{\hat q}^{\mu_{1}}\ldots{\hat q}^{\mu_{2n+1}}&=&
 \sum_{k=0}^{n} \frac{c'_{n,k}}{(2n+1)!}
 \Big[{\hat U}_{2k+1}^{\mu_{1}\ldots\mu_{2k+1}}({\hat q})
 \delta^{\mu_{2k+2}\mu_{2k+3}}\ldots \delta^{\mu_{2n}\mu_{2n+1}}
 \nonumber \\ && + {\hbox{ all symmetric permutations in }}
 (\mu_{1}\ldots\mu_{2n+1})\Big]
 \,. \label{418}
\end{eqnarray}
So all ${\hat q}^{\mu_{1}}\ldots{\hat q}^{\mu_{n}}$ can be expanded in
terms of our generalized Chebyshev polynomials. Since any smooth
function of $\vec{q}$ can always be expanded in terms of ${\hat
q}^{\mu_{1}}\ldots{\hat q}^{\mu_{n}}$ times a function which depends
only on the absolute value $q$, this means that we can always expand
our four-point functions in terms of these Chebyshev functions.

%%%%%%%%%%%%%%%%%%%%%%%%%%%%%%%%%%%%%%%%%%%%%%%%%%%%%%%%%%%%%%%%%%%%%%%%%%%%%

%%%%%%%%%%%%%%%%%%%%%%%%%%%%%%%%%%%%%%%%%%%%%%%%%%%%%%%%%%%%%%%%%%%%%%%%%%%%%

%%%%%%%%%%%%%%%%%%%%%%%%%%%%%%%%%%%%%%%%%%%%%%%%%%%%%%%%%%%%%%%%%%%%%%%%%%%%%
\end{document}